\documentclass[fleqn,usenatbib,twocolumn]{mnras}

\usepackage{newtxtext,newtxmath}
\usepackage{xspace}

\usepackage[T1]{fontenc}

\DeclareRobustCommand{\VAN}[3]{#2}
\let\VANthebibliography\thebibliography
\def\thebibliography{\DeclareRobustCommand{\VAN}[3]{##3}\VANthebibliography}

\usepackage{graphicx}	
\usepackage{amsmath}	
\usepackage{gensymb}
\usepackage{subcaption}
\usepackage{orcidlink}
\usepackage{booktabs}
\usepackage{siunitx}
\usepackage{anyfontsize} 
\usepackage[table]{xcolor}
\usepackage{upgreek}
\usepackage{makecell}
\usepackage{placeins}

\newcommand{\rone}{FRB~20121102A\xspace}

\newcommand{\sgr}{SGR~1935$+$2154\xspace}
\newcommand{\rsixseven}{FRB~20201124A\xspace}
\newcommand{\roneoneseven}{FRB~20220912A\xspace}
\newcommand{\ronefourseven}{FRB~20240114A\xspace}
\newcommand{\ronezeronine}{FRB~20220529A\xspace}

\newcommand{\rthree}{FRB~20180916B\xspace}

\newcommand{\rmkt}{FRB~20240619D\xspace}

\newcommand{\sfxc}{{\tt SFXC}\xspace}

\newcommand{\digifil}{{\tt digifil}\xspace}

\def\torun{Toru\'n\xspace}
\def\nancay{Nan\c{c}ay\xspace}
\newcommand{\dmunit}{pc\,cm$^{-3}$\xspace}
\newcommand{\dmunityr}{pc\,cm$^{-3}$\,yr$^{-1}$\xspace}
\newcommand{\rmunit}{rad\,m$^{-2}$\xspace}
\newcommand{\microsec}{$\upmu \mathrm{s}$\xspace}
\newcommand{\erghz}{erg\,Hz$^{-1}$\xspace}
\newcommand{\erghzs}{erg\,s$^{-1}$\,Hz$^{-1}$\xspace}
\newcommand{\eclat}{\'ECLAT\xspace}
\newcommand{\rom}[1]{\uppercase\expandafter{\romannumeral #1\relax}}

\makeatother
\usepackage[switch]{lineno} 

\newcommand{\vlbilocal}{$\alpha=21^{\mathrm{h}}27^{\mathrm{m}}39.835^{\mathrm{s}}$, $\delta = +4\degr19\arcmin45.668\arcsec$ (J2000; ICRF)}

\title[\ronefourseven: energetics]{A 4200-hour HyperFlash and \eclat campaign on the hyperactive \ronefourseven: constraining energetics with the most brilliant bursts}

\author[O.S. Ould-Boukattine et al.]
{O.~S.~Ould-Boukattine \orcidlink{0000-0001-9381-8466}$^{1,2}$\thanks{E-mail: ouldboukattine@astron.nl},
A.~J.~Cooper \orcidlink{0000-0002-4033-3139}$^{3}$,
A.~M.~Cook \orcidlink{0000-0001-6422-8125}$^{4,5,2}$,
J.~W.~T.~Hessels \orcidlink{0000-0003-2317-1446}$^{4,5,2,1}$, 
D.~M.~Hewitt \orcidlink{0000-0002-5794-2360}$^{2}$, \newauthor
J.~Huang \orcidlink{0000-0002-8043-0048}$^{4,5}$, 
I.~Cognard \orcidlink{0000-0002-1775-9692}$^{6,7}$, 
T.~J.~Dijkema \orcidlink{0000-0001-7551-4493}$^{1,8}$, 
M.~P.~Gawro\'nski \orcidlink{0000-0003-4056-4903}$^{9}$,
W.~Herrmann \orcidlink{0000-0001-5806-446X}$^{10}$, 
F.~Kirsten \orcidlink{0000-0001-6664-8668}$^{11,1}$, \newauthor
A.~Moroianu \orcidlink{0000-0003-1936-9062}$^{2}$, 
Z.~Pleunis \orcidlink{0000-0002-4795-697X}$^{2,1}$,
W.~Puchalska \orcidlink{0000-0003-2422-6605}$^{9}$,
S.~Ranguin \orcidlink{0009-0004-1397-9331}$^{2}$,
M.~P.~Snelders\orcidlink{0000-0001-6170-2282}$^{1,2}$, 
and T.~Telkamp \orcidlink{0009-0002-5048-2573}$^{8}$ \newauthor
\\
$^{1}$ASTRON, Netherlands Institute for Radio Astronomy, Oude Hoogeveensedijk 4, 7991 PD Dwingeloo, The Netherlands\\
$^{2}$Anton Pannekoek Institute for Astronomy, University of Amsterdam, Science Park 904, 1098 XH, Amsterdam, The Netherlands\\
$^{3}$Astrophysics, The University of Oxford, Keble Road, Oxford, OX1 3RH, UK\\
$^{4}$Trottier Space Institute, McGill University, 3550 rue University, Montr\'eal, QC H3A~2A7, Canada\\
$^{5}$Department of Physics, McGill University, 3600 rue University, Montr\'eal, QC H3A~2T8, Canada\\
$^{6}$Station de Radioastronomie de Nan\c{c}ay, Observatoire de Paris, PSL University, CNRS, Universit\'{e} d'Orl\'{e}ans, F-18330 Nan\c{c}ay, France\\
$^{7}$Laboratoire de Physique et Chimie de l'Environnement et de l'Espace LPC2E UMR7328, Universit\'{e} d'Orl\'{e}ans, CNRS, F-45071 Orl\'{e}ans, France\\
$^{8}$CAMRAS Dwingeloo Radio Telescope Foundation, Oude Hoogeveensedijk 4, 7991 PD Dwingeloo, The Netherlands \\
$^{9}$Institute of Astronomy, Faculty of Physics, Astronomy and Informatics, Nicolaus Copernicus University, Grudziadzka 5, 87-100 \torun, Poland\\
$^{10}$Astropeiler Stockert e.V., Astropeiler 1-4, 53902 Bad M\"{u}nstereifel, Germany\\
$^{11}$Department of Physics and Astronomy, Chalmers University of Technology, Onsala Space Observatory, 439 92, Onsala, Sweden\\
}

\date{Accepted 2026 October 12. Received 2026 October 12; in original form 2026 May 19}

\pubyear{\the\year{}}

\begin{document}
\label{firstpage}
\pagerange{\pageref{firstpage}--\pageref{lastpage}}
\maketitle

\begin{abstract}
Hyperactive repeaters provide a unique window into the evolving environments and energy budgets of fast radio burst (FRB) sources, though they may not be representative of the FRB population in general. High-cadence observations are key to capturing the rarest and most energetic bursts, which occur only once per hundreds to thousands of hours. Here we present an unprecedented $4{,}200$-hour observing campaign targeting \ronefourseven as part of the HyperFlash and \eclat FRB monitoring programs. Over $806$~days, we detected $178$ high-energy ($\sim$$10^{40-42}$\,erg) bursts with HyperFlash, which together amount to $4.4 \times 10^{42}$\,erg of released radio energy (assuming isotropic emission and 1-GHz emission bandwidth). The cumulative energy of the HyperFlash bursts is about twice that of $\sim$$11{,}000$ lower-energy bursts detected with FAST, emphasising the significant role that the highest-energy bursts play in depleting the central engine's stored energy. In fact, the single most brilliant burst from our sample, which we term the STROOP, contributes roughly $1/3$ of all the energy we measure, and is at the maximum energy seen in studies of both repeating and apparently one-off FRBs alike. We also find a break in the burst energy distribution at $\sim$$2\times10^{40}$\,erg and a linear dispersion measure (DM) increase of $+0.96 \pm 0.06$\,\dmunit over a period of $318$\,days. We discuss these findings in the context of a magnetar source model and highlight comparisons with the energetics of intermediate and giant X-ray/$\gamma$-ray flares from Galactic sources.
\end{abstract}


\clearpage

\section{Introduction}

Of the already small fraction ($2.4\%$) of fast radio burst (FRB) sources that are known to repeat \citep[][]{spitler_2016_natur,cook_2026_arxiv}, an even smaller handful have been observed to be `hyperactive', producing hundreds to thousands of observed bursts over weeks to months \citep[e.g.,][]{konijn_2024_mnras}. Within the broader FRB population, such sources are particularly useful for constraining source environments \citep[e.g.,][]{michilli_2018_natur} and energetics \citep[e.g.,][]{ouldboukattine_2026_mnras_2} --- even though it remains unclear whether all FRBs share a common progenitor \citep[e.g.,][and references therein]{petroff_2022_aarv}. 

\ronefourseven was discovered and identified as a hyperactive repeater by CHIME/FRB \citep{shin_2024_atel, shin_2026_apj}. The FRB source was subsequently localised to arcsecond precision using MeerKAT \citep{tian_2024_mnras}. Observations with the European VLBI Network (EVN) by \citet{snelders_2024_atel}, as part of the PRECISE project \citep{marcote_precise}, further improved the localisation to milliarcsecond precision at \vlbilocal. This enabled the identification of the host as a satellite galaxy of a more massive galaxy, at a redshift of $z=0.130287$, corresponding to a luminosity distance of $616$\,Mpc \citep{bhardwaj_2025_apjl}. \ronefourseven has revealed itself as one of the most hyperactive repeating sources to date, with thousands of detections reported from, e.g., the Five-hundred-meter Aperture Spherical Telescope \citep[FAST;][]{zhang_2025_arxiv}, the \nancay Radio Telescope \citep[NRT;][]{hewitt_2024_atel}, and the upgraded Giant Metrewave Radio Telescope \citep[uGMRT;][]{panda_2025_apj}. The source has been detected across a wide range of radio frequencies, from as low as $300$\,MHz \citep{ouldboukattine_2024_atel_16432} up to $6$\,GHz \citep{limaye_2026_aa}, while remaining undetected at $8.6$\,GHz \citep{wang_2025_apj}. \ronefourseven remained active for $\sim$$16$\,months  from its discovery until May 2025, with intermittent non-detection phases and apparent episodes of enhanced activity \citep{ouldboukattine_2025_atel_16967}.

The identification of hyperactive repeaters has led some authors to suggest that magnetars, the leading candidate sources to power (repeating) FRBs, may exhaust their external magnetospheric energy reservoirs ($E_{\rm mag} \sim 10^{47} (B_{\rm s}/10^{15} {\rm G})^2 \, {\rm erg}$) in just a few weeks \citep{zhang_2025_arxiv,ouldboukattine_2026_mnras_2}. This is sometimes referred to in the literature as the energy budget crisis. For an FRB source that emits bursts in all directions, but where each burst has a beaming angle $0 < f_{\rm b} \leq 1$, the increase in observed energy due to beaming $E_{\rm obs} = E_{\rm true}/f_{\rm b}$ is offset by the bursts missed by the observer $N_{\rm obs} = f_{\rm b} N_{\rm true}$. This means that the total energy observed may reflect the true global energetics: $E_{\rm total, obs} \approx  E_{\rm total, true}$. Assuming a radio efficiency factor of $\eta_{\rm r} \ll 1$ where $E_{\rm radio} = \eta_{\rm r} E_{\rm total}$, the total energy produced by these hyperactive repeaters is very large. However, this order of magnitude estimate for the total emitted energy can be significantly relaxed by considering more efficient radio emission $\eta_{\rm r} \gtrsim 10^{-4}$ or the replenishment of the external magnetic energy via core field expulsion (e.g., \citealt{lander_26}). Nevertheless, these considerations have led some authors to consider whether hyperactive repeaters like \ronefourseven may represent a special geometric case. For example, \cite{zhang_hu_2025} present a model in which repeating sources of FRBs are magnetars in binaries where the magnetar's rotation and magnetic axes are aligned (see also \citealt{beniamini_kumar_2025}). Moreover, for the most active repeating sources, these axes may be further aligned with the orbital axis. Similarly, \cite{Luo_2025_starquake} propose a scenario in which prolific repeaters represent isolated magnetars undergoing starquake activity with small magnetic inclination angles, providing a similarly privileged observing position. If the observer is well-aligned with these axes along which FRBs are preferentially produced and beamed, then energy budget concerns are significantly alleviated as $E_{\rm total, obs} \gg  E_{\rm total, true}$. This is because fewer FRBs are missed by the observer, meaning the inferred (isotropic-equivalent) energy can exceed the true underlying energy. Moreover, aligned geometries ease further concerns regarding the non-detection of a spin periodicity amongst thousands of FRBs.  

Here we report high-cadence monitoring of \ronefourseven as part of the HyperFlash and \eclat observing campaigns, comprising more than $4200$\,h of observations collected over more than $800$\,days. We detected $178$ high-fluence bursts ($>8$\,Jy\,ms), of which $41$ were co-detected by multiple telescopes, for a total of $219$ detections. In this paper, we focus on \ronefourseven's energetics and time-variable propagation effects, including the cumulative burst energy distribution, cumulative energy release, and time-evolving DM.

\section{Observations \& burst search}

\subsection{HyperFlash \& \eclat}
We observed \ronefourseven using $5$ European radio telescopes as part of the HyperFlash observational program (PI: O.~S.~Ould-Boukattine) between 2024 January 29 (MJD~$60338$) and 2026 April 13 (MJD~$61143$). HyperFlash is a high-cadence FRB monitoring program in which $25$–$32$\,m class radio telescopes aim to observe the brightest and rarest FRBs at complementary radio wavelengths. Observations are primarily conducted at L-band ($\sim$$1.4$\,GHz), as all participating telescopes operate at this wavelength, with additional coverage at P-band ($324$\,MHz) and C-band ($4.8$\,GHz). The participating stations are the 25-m Westerbork dish RT-1 (Wb; the Netherlands), the 25-m Onsala telescope (O8; Sweden), the 32-m \torun telescope (Tr; Poland), the 25-m Stockert telescope (St; Germany), and the 25-m Dwingeloo telescope (Dw; the Netherlands).

HyperFlash observations targetting \ronefourseven were initiated on 2024 January 25, after identifying a repeating source in the CHIME/FRB VOEvent data stream \citep{chime_2018_apj,abbott_2025_aj}. Our initial pointing was based on the average position of the first reported bursts in the data stream. Following the initial report of the source in The Astronomer’s Telegram \citep[ATel;][]{shin_2024_atel}, we found that our pointing was offset by $1.24$\arcmin\ from the more accurately constrained CHIME/FRB baseband localisation reported therein. We note and caution that localisations reported in the CHIME/FRB VOEvent stream are preliminary and may differ from refined positions derived from the baseband data. We subsequently updated our pointing, first to the CHIME/FRB ATel localisation reported on 2024 January 29, then to the arcsecond localisation from MeerKAT \citep{tian_2024_mnras}, and finally on 2024 March 20 to the milliarcsecond localisation reported by \citet{snelders_2024_atel}: \vlbilocal. The small offsets between these reported localisations are all well within the full width at half maximum (FWHM) of the primary beams of the respective telescopes, which vary across the observed frequency bands from $0.1$\degree\ (C-band) to $2.3$\degree\ (P-band). 

In addition to the HyperFlash dataset, we include lower-cadence observations from the \nancay Radio Telescope (NRT; France). These observations were carried out as part of the Extragalactic Coherent Light from Astrophysical Transients (\eclat) programme (PI: D.~M.~Hewitt). In this paper we only include the burst fluences from $673$~detections at L-band between 2024 February 4 (MJD~$60344$) and 2025 February 17 (MJD~$60723$). A full analysis of the NRT burst sample, including bursts detected at S-band ($\sim$$2.1$\,GHz), is currently forthcoming (J.~Huang~in prep.). The NRT is a Kraus-type reflecting radio telescope with a collecting area comparable to that of a $\sim$$95$-m diameter parabolic dish. Since the start of 2022, the \eclat programme has been monitoring a sample of $\sim$\,$20$ repeating FRB sources, with each source observed for approximately $1$\,hr\,week$^{-1}$. 

Over the $806$\,days spanned by the HyperFlash and \eclat campaigns, we accumulated $4{,}233.58$\,hours of observing --- making this, to our knowledge, the longest dedicated campaign on a single repeating FRB source to date. Over the entire observing campaign, HyperFlash detected $178$~bursts, of which $41$ were simultaneously detected by multiple telescopes; these simultaneous detections were all at L-band. All burst properties are listed in Appendix~Table~\ref{tab:hyperflash_burst_prop_appendix} and are available in \texttt{.csv} format as part of the Supplementary Material. An overview of the campaign, including burst detections, is shown in Appendix Figure~\ref{fig:obs_campaign_r147}. Details of the observations, including detections and completeness thresholds as well as the time observed per telescope, are provided in Appendix Table~\ref{tab:obs_coverage}.

\subsubsection{Westerbork, Onsala \& \torun}
The radio telescopes at Westerbork, Onsala, and \torun use the same FRB search pipeline, as they are all part of the European VLBI Network (EVN) and employ similar recording setups. The search pipeline, \texttt{FRB-baseband}, has been described previously; most recently, in \citet{ouldboukattine_2026_MNRAS}.

We record the raw voltage data (`baseband' or `waveform data') in \texttt{VDIF} format with dual-polarization channels and 2-bit sampling \citep{whitney_2010_ivs}. The voltage data is then processed into 8-bit total-intensity \texttt{SIGPROC} filterbank files using \texttt{digifil} \citep{vanstraten_2011_pasa}. To search the data and correct for dispersion, we generate filterbank files with time and frequency resolutions optimized for each observing frequency, balancing sensitivity and computational cost: $128$\,\microsec and $7.8125$\,kHz at P-band ($324$\,MHz), $16-64$\,\microsec and $15.625$\,kHz at L-band ($\sim$$1.4$\,GHz), and $16$\,\microsec and $250$\,kHz at C-band ($\sim$$4.8$\,GHz). We mitigate radio frequency interference (RFI) across the different frequency setups and channelisations using static masks for frequency channels known to be affected by RFI. 
We search the data for burst candidates using \texttt{Heimdall} \citep{barsdell_2012_mnras}, adopting a signal-to-noise (S/N) threshold of $7$ and a maximum boxcar width of $1024$\,samples, corresponding to $16.384$\,–\,$131.072$\,ms depending on the time resolution. We further enable the \texttt{boxcar-renorm} option to revert to the original boxcar renormalisation calculation of the filtered time series (see Appendix~Section~\ref{sec:heimdall_option} for further discussion). We limit the DM search range to $\pm$\,$50$\,\dmunit around the discovery DM of $527.7$\,\dmunit \citep{shin_2024_atel}. Burst candidates are subsequently classified using \texttt{FETCH} \citep{agarwal_2020_mnras}, specifically employing models A and H. These models were selected based on extensive testing on previous datasets, which demonstrated their reliability in minimizing false positive rates while also remaining accurate \citep{snelders_2020}. As an additional check, we manually inspect the burst candidates found within $\pm$\,$5$\,\dmunit of the known DM of the source.  
Over the entire observing campaign, Westerbork detected $62$~bursts, $60$ at L-band and $2$ at P-band. One of the P-band bursts was also detected by the Northern Cross radio telescope \citep{pelliciari_2024_atel}. Onsala detected $48$~bursts and \torun detected $37$~bursts. An overview of the observations is given in Appendix Table~\ref{tab:obs_coverage}.

\subsubsection{Stockert}
The recording and search pipeline employed at Stockert has most recently been described in \citet{ouldboukattine_2026_MNRAS}. We record 32-bit total intensity data using the Pulsar Fast Fourier Transform Spectrometer (PFFTS) backend \citep{barr_2013_mnras}, which is subsequently converted into a 32-bit float filterbank format using the \texttt{filterbank} tool from the \texttt{SIGPROC} software suite. The highest attainable resolution of the data is a time resolution of $218.45$\,\microsec and a frequency resolution of $586$\,kHz. To search the data for bursts, we make use of tools from the \texttt{PRESTO} software suite \citep{ransom_2011_ascl}. The data are first dedispersed to $527.7$\,\dmunit using \texttt{prepsubband}, while RFI is mitigated with \texttt{rfifind}. We then identify burst candidates using \texttt{single\_pulse\_search.py}, adopting a S/N threshold of $8$ and a maximum boxcar width of $300$, corresponding to approximately $65$\,ms. All candidates are subsequently classified with \texttt{FETCH}, where we employ models A and H, and apply a threshold of $>50$\,\%. Stockert detected $69$~bursts; see Appendix Table~\ref{tab:obs_coverage} for a summary.

\subsubsection{Dwingeloo}
Observations at Dwingeloo were conducted using a custom-built pipeline, \texttt{Dwingeloo-FRBscripts}, capable of simultaneously recording data in both P-band ($400-420$\,MHz) at 12-bit and L-band ($1.2-1.4$\,GHz) at 14-bit. Filterbank and baseband data are recorded and written to disk using \texttt{vrt-iq-tools}. The baseband data are stored in 10-minute temporary buffers; extracts from the data are stored when the semi-real-time pipeline detects an event. We record 32-bit float filterbank data; at P-band dual-polarization (linear) with a time resolution of $192$\,\microsec and a frequency resolution of $125$\,kHz. At L-band, the data are recorded with a single polarization with a time resolution of $198.4$\,\microsec and a frequency resolution of $31.25$\,kHz.

The data are processed using the \texttt{PRESTO} software suite \citep{ransom_2011_ascl}. We generate dedispersion plans with \texttt{DDplan.py} and mitigate RFI using \texttt{rfifind}. Burst candidates are identified using \texttt{single\_pulse\_search.py} for which we set a S/N threshold of $6$. Burst candidate events are subsequently clustered in DM-time using the \texttt{DBSCAN} algorithm \citep{db_scan_1996}. For classification, we employ FETCH, specifically utilizing model A with a detection threshold of 50\% \citep{agarwal_2020_mnras}.

The feed of the Dwingeloo telescope was optimized for amateur radio observations at 1296 MHz, while the telescope itself is located in a strong RFI environment. Therefore, each of the $3$~bursts reported from Dwingeloo in this work were only confirmed after independent detections at Westerbork, \torun, Stockert, or Onsala.

\subsubsection{\nancay Radio Telescope (\eclat)} \label{nrt_section22}
The recording setup and burst search strategy deployed at the NRT have been described in detail by \cite{hewitt_2023_mnras} and \citet{konijn_2024_mnras}. We conducted observations using the low-frequency receiver ($1.1-1.8$\,GHz) of the focal plane instrument {\it Foyer Optimisé pour le Radio Télescope}. We record $512$\,MHz of bandwidth centered at $1484$\,MHz, divided into $128$ frequency channels of width $4$\,MHz. At $1.4$\,GHz, the system has a temperature of $\textrm{T}_\textrm{sys} \approx 35$\,K and a gain of $\textrm{G} \approx 1.4$\,K\,Jy$^{-1}$. To mitigate intra-channel dispersive smearing, we applied online coherent dedispersion, adopting a DM of $527.7$\,\dmunit for observations targeting \ronefourseven, the DM reported in the discovery ATel \citep{shin_2024_atel}. The NRT campaign reported in this study includes observations between 2024 February 4 (MJD~$60344$) and 2025 February 17 (MJD~$60723$), during which a total of $673$ bursts were detected at L-band over $55.99$\,hours (see Appendix Table~\ref{tab:obs_coverage}). 

\section{Analysis \& results}

\subsection{Burst properties}

\subsubsection{HyperFlash}
Burst properties are measured using \texttt{filterbank} data created with a custom workflow previously used and described in \cite{ouldboukattine_2026_MNRAS}. At Westerbork, Onsala, and \torun, digitisation artefacts arise due to the limited dynamic range of 2-bit sampling combined with the brightness of the bursts; this effect scales with the brightness occupancy within a single subband (see Appendix Table~\ref{tab:obs_coverage} for the subband widths). A detailed description of the correction of these digitisation effects is provided in Appendix~B2 of \cite{ouldboukattine_2026_mnras_2}, and an illustrative example is shown in Extended Figure~4 of \cite{kirsten_2024_natas}. In essence, when baseband data is available, we apply the scattered power correction \citep[SPC;][]{vanstraten_2013_apjs} to bursts, using \texttt{digifil}.

We adopt a single DM of $527.7$\,\dmunit\ for the burst property analysis. Although our data show that the DM increases over time (see Section~\ref{sec:variable_dm}), the maximum increase is only $<0.2\%$, which does not significantly affect the fluence measurements. RFI is mitigated using \texttt{psrzap} and \texttt{pazi} from the \texttt{PSRCHIVE} software suite, and the top and bottom channels of each subband are additionally flagged. The start and stop times of the bursts are determined manually. To measure the frequency extent of each burst, we compute the two-dimensional auto-correlation function (ACF) and fit a Gaussian function along the frequency axis; if twice the full width at half maximum (FWHM) exceeds $75$\,\% of the total observing bandwidth, the burst is considered to span the full bandwidth. We then sum over this frequency extent and convert the resulting time series to flux densities using the radiometer equation to determine the fluence, assuming a constant system equivalent flux density (SEFD) for all observations (see Appendix Table~\ref{tab:obs_coverage}). Finally, the time of arrival (ToA) is determined by fitting a Gaussian to the burst time profile, with the ToA defined at the center of the fit. These ToAs are referenced to barycentric arrival times in the Barycentric Dynamical Timescale (TDB), assuming a dispersion constant of $\mathcal{D} = 1/(2.41 \times 10^{-4})$\,MHz$^{2}$\,pc$^{-1}$\,cm$^{3}$\,s relative to infinite frequency. The locations of the HyperFlash telescopes in an Earth-centered, Earth-fixed (X,Y,Z) coordinate system are given in Appendix Table~\ref{tab:hyperflash_burst_prop_appendix}.

\subsubsection{\eclat}
We made use of a bespoke burst property analysis pipeline, \texttt{ECLAT-burst-analyzer}, previously described in \cite{ouldboukattine_2026_MNRAS} and to be presented in more detail in an upcoming paper on the NRT burst sample (J.~Huang~in prep.). In brief, the pipeline dedisperses each burst incoherently to a fixed DM of $527.7$\,\dmunit. The bandpass is corrected by subtracting the mean and normalizing by the standard deviation per channel. RFI is mitigated using \texttt{jess}\footnote{\url{https://github.com/josephwkania/jess}} as a first pass, where frequency channels with skewness outliers exceeding $3\sigma$ are flagged, followed by additional manual flagging of any remaining contamination. We manually determine both the temporal duration and spectral extent of each burst. The fluence is then calculated by summing over the full bandwidth, integrating over the burst duration, and applying the radiometer equation.

\subsection{Cumulative sum of isotropic energies}
\begin{figure*}
    \centering
    \includegraphics[width=\linewidth]{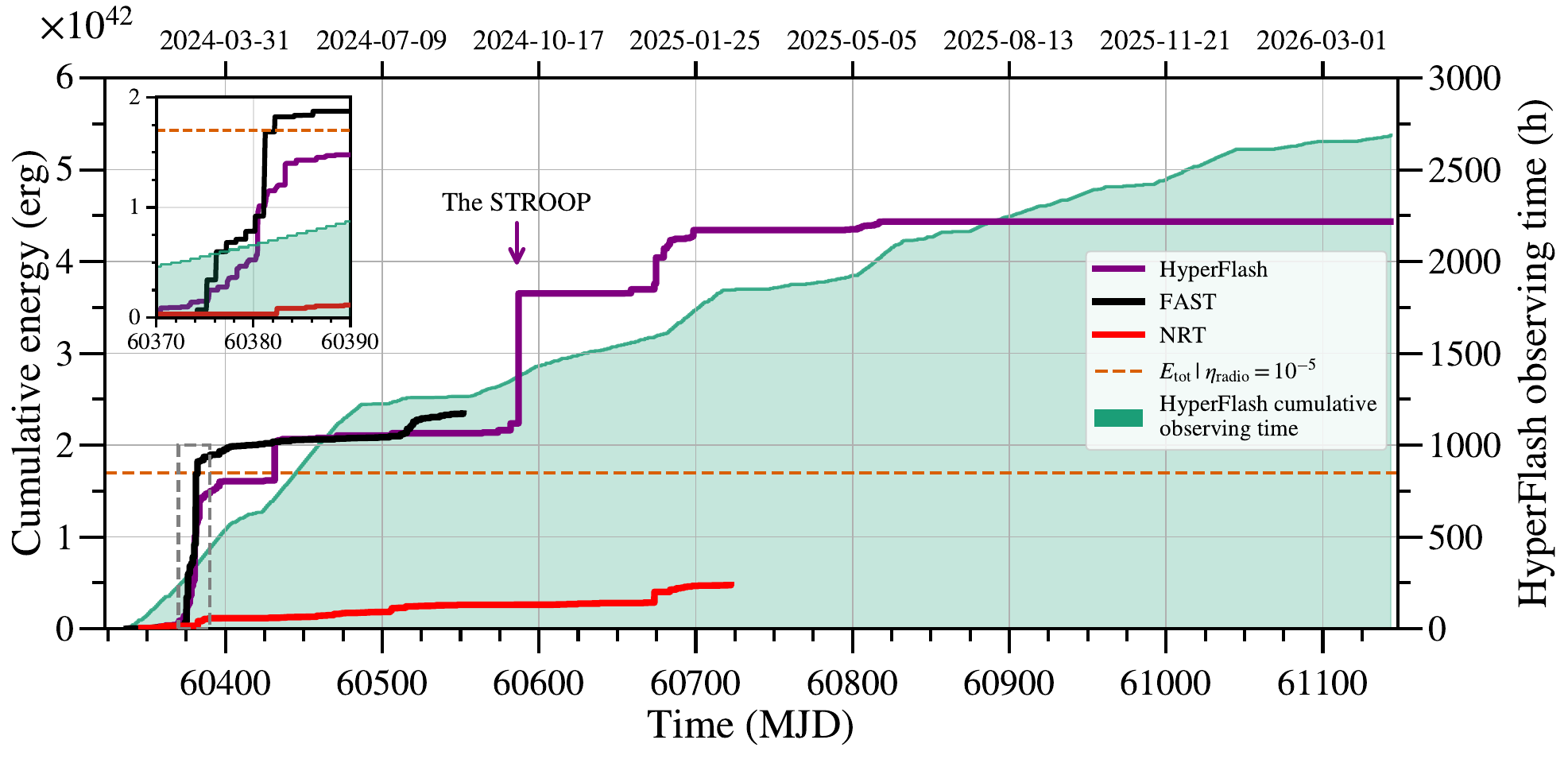}
    \caption{Cumulative sum of isotropic-equivalent burst energies for the HyperFlash sample (purple), the FAST burst sample (black: 34\,h total observing time; \citealt{zhang_2025_arxiv}) and the NRT burst sample (red: 56\,h of observations; J.~Huang~in prep.)). The energies have been scaled to a $1$\,GHz emission bandwidth. The corresponding cumulative emitted energies, as observed by these telescopes, are $E^{\mathrm{HF}}_{\mathrm{tot}} = 4.4 \times 10^{42}$\,erg,  $E^{\mathrm{FAST}}_{\mathrm{tot}} = 2.3 \times 10^{42}$\,erg and $E^{\mathrm{NRT}}_{\mathrm{tot}} =4.8 \times 10^{41}$\,erg. The brightest event in the HyperFlash sample, shown by an arrow at MJD~$60592$, ``the STROOP'' (Figure~\ref{fig:the_stroop}), has an inferred isotropic-equivalent energy of $E_{\mathrm{STROOP}} = 1.4 \times 10^{42}$\,erg. The green shaded curve represents the unique, cumulative observing time of the HyperFlash campaign at L-band ($\sim$$1.4$\,GHz), for a total of $2{,}688.24$\,hours. The orange-dotted line indicates the estimated external magnetic energy of a magnetar, assuming a dipolar field and a radio efficiency of $\eta_{\mathrm{r}} = 10^{-5}$, at $1.7\times10^{42}$\,erg. The inset in the top left zooms in on a $10$-day period of exceptionally high burst activity.}
    \label{fig:cumulative_sum_energy_hf_nrt_fast}
\end{figure*}

In Figure~\ref{fig:cumulative_sum_energy_hf_nrt_fast} we present the cumulative sum of isotropic-equivalent burst energies detected at L-band ($\sim$$1.4$\,GHz), plotted against the HyperFlash cumulative observing time. Fluences from the different observing campaigns are converted to isotropic-equivalent energies and scaled to a canonical $1$\,GHz emission bandwidth. This is standard practice in the literature because the beaming fraction and total emission bandwidth are unknown. The HyperFlash sample of bursts detected at L-band (purple solid line) comprises $176$ bursts; in cases of multiple detections of the same burst, we retain the highest-energy detection. The total isotropic-equivalent energy of all HyperFlash-detected bursts is $E^{\mathrm{HF}}_{\mathrm{tot}} = 4.4 \times 10^{42}$\,erg. We additionally show the cumulative sums for FAST \citep{zhang_2025_arxiv} and NRT burst samples, with total energies of $E^{\mathrm{FAST}}_{\mathrm{tot}} = 2.3 \times 10^{42}$\,erg and $E^{\mathrm{NRT}}_{\mathrm{tot}} = 4.8 \times 10^{41}$\,erg, respectively. Note that the NRT campaign unfortunately did not cover the high-activity window of \ronefourseven in 2024 March due to maintenance of the telescope. The horizontal orange-dotted line is an estimate of the energy available for radio emission. Assuming a dipolar magnetic field, a crude estimate of the external magnetospheric energy reservoir of a magnetar ($U_{\rm B}$) is given by
\begin{equation} \label{eq:total_mag_energy}
U_{\rm B} \approx \frac{R_{\rm NS}^3 B_{\rm s}^{2}}{6} \approx 1.7 \times 10^{47} \, {\rm erg} \; \bigg(\frac{R_{\rm NS}}{10^{6} \, {\rm cm}}\bigg)^3 \, \bigg(\frac{B_{\rm s}}{10^{15} \, {\rm G}}\bigg)^2
\end{equation}
where $R_{\rm NS}$ is the neutron star radius and $B_{\rm s}$ is the surface magnetic field. Assuming a radio efficiency of $\eta_{\rm r} \sim 10^{-5}$, this corresponds to an available radio energy of $\eta_{\rm r} U_{\rm B} \approx 1.7 \times 10^{42}$\,erg.

The cumulative observing time accounts for overlap between the five HyperFlash stations and NRT observing at L-band and reaches a total of $2{,}688.24$\,hours over the full campaign. Additionally, Appendix~Figure~\ref{fig:cumulative_sum_energy_hf} shows the cumulative sum of isotropic-equivalent energies as a function of daily observing time. This plot demonstrates that, at the start of the campaign, observations were carried out daily over the full visibility window of the source in Europe ($\sim$$11.2$\,hours), and highlights the overall consistency of the observing cadence throughout the campaign.

\begin{figure*}
    \centering
    \includegraphics[width=\linewidth]{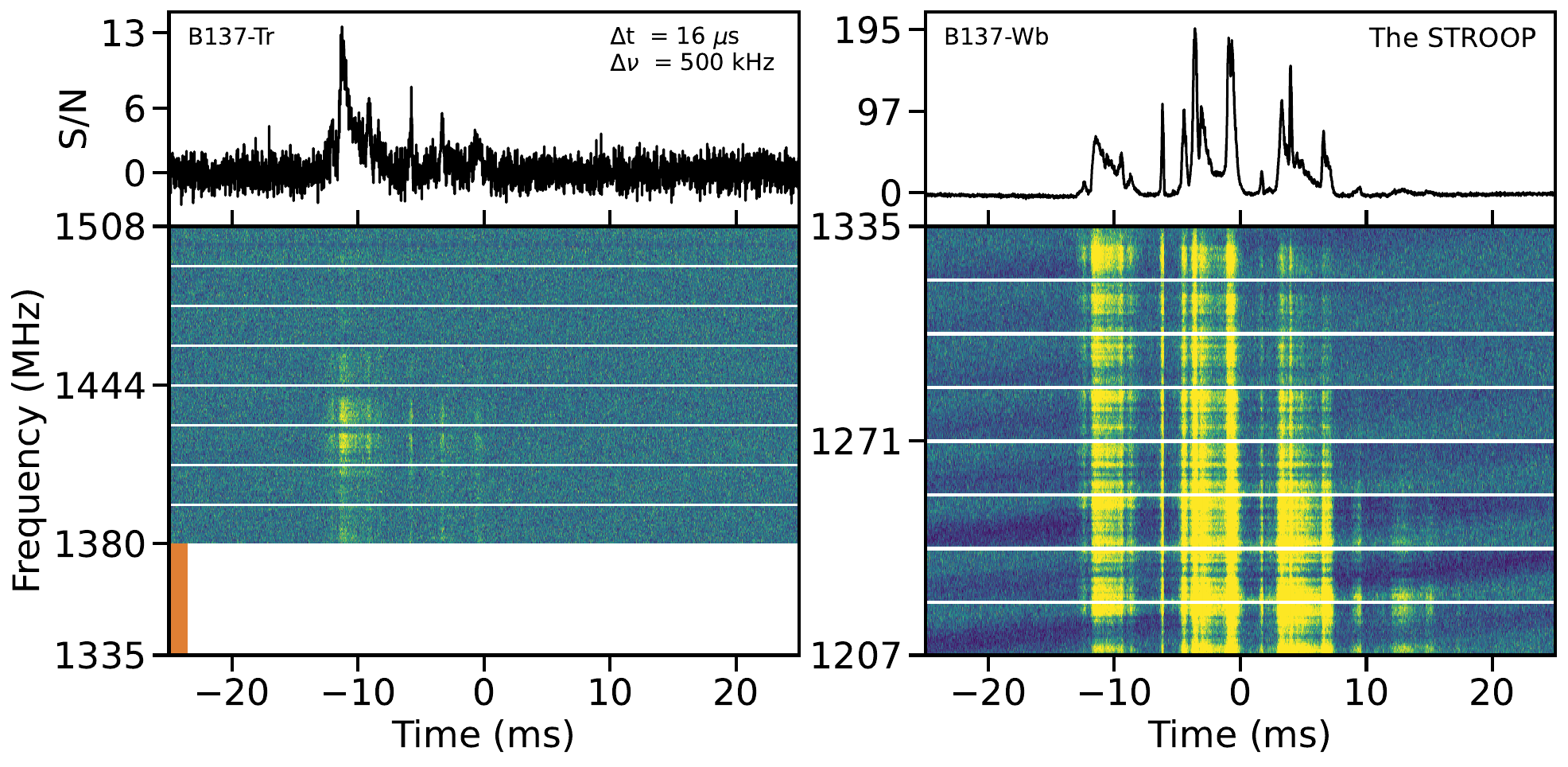}
    \caption{
    Dynamic spectra and time series of the brightest burst in our sample, which we have named ``the STROOP'' (\textit{in Dutch:} ``de \textbf{ST}erkste snelle \textbf{R}adioflits \textbf{O}oit \textbf{OP}gevangen''). The STROOP was detected with both \torun (left) and Westerbork (right). The orange bar indicates the gap in observing range of $45$\,MHz between \torun and Westerbork. In both panels, the time resolution is $16$\,\microsec and the frequency resolution is $500$\,kHz. The blue diamond patterns visible in the Westerbork dynamic spectra are digitisation artefacts due to limited dynamic range, and scale with the signal intensity within each subband. The masked white channels represent the subband edges, which have been flagged. Both bursts have been dedispersed coherently to $528.468$\,\dmunit, which optimises the S/N of the narrowest and brightest components (Section~\ref{sec:variable_dm}). We measure a fluence for the Westerbork detection of the STROOP of $3{,}983$\,Jy\,ms and, assuming a canonical bandwidth of $1$\,GHz, the isotropic-equivalent energy is $E_{\mathrm{STROOP}} = 1.4 \times 10^{42}$\,erg; see Section~\ref{sec:disc_energy_stroop} for a discussion of the energy implications.  
    }
    \label{fig:the_stroop}
\end{figure*}

We identify an exceptionally bright burst in the HyperFlash sample, which we name ``the STROOP'' (\textit{in Dutch:} ``de \textbf{ST}erkste snelle \textbf{R}adioflits \textbf{O}oit \textbf{OP}gevangen''\footnote{In English: ``the most powerful FRB ever caught''.}; see Figure~\ref{fig:the_stroop}), with an inferred isotropic energy of $E_{\mathrm{STROOP}} = 1.4 \times 10^{42}$\,erg. This corresponds to $61$\,\% of the total energy observed in the FAST sample, or equivalently the cumulative contribution of the first $10{,}951$ bursts when ordered by energy from lowest to highest. The STROOP alone accounts for $32$\,\% of the total cumulative energy of the HyperFlash sample. This demonstrates the importance of detecting the rarest bursts in order to gain a full understanding of the global energetics of repeating FRBs, and of high-cadence, low-sensitivity observations.

\subsection{Cumulative burst rates} \label{sec:burstrate_how}
Cumulative burst energy distributions are commonly described by a power-law above a completeness limit, defined as the energy threshold above which all bursts are detected (e.g., \citealt{hewitt_2022_mnras}). Below this threshold, the distribution typically exhibits a turnover toward lower energies, generally attributed to incompleteness from sensitivity limitations \citep[e.g.,][]{gourdji_2019_apjl} --- although this turnover at lower energies has also been interpreted to be astrophysical in origin in studies conducted with FAST \citep[e.g.,][]{li_2021_natur}. The cumulative distribution follows $R(>E_{\nu}) \propto E_{\nu}^{\gamma}$, where $R$ is the burst rate (typically in bursts per hour), $E_{\nu}$ is the spectral energy density (\erghz), and $\gamma$ is the cumulative power-law index. We express burst energies as spectral energy densities, $E_{\nu} = E / \nu$, to enable consistent comparison between different telescopes, where $\nu$ represents the spectral extent of the burst. We convert fluences to spectral energy density via \citep{macquart_2018_mnras}:
\begin{equation}
E_{\nu} = \frac{\mathcal{F} \, 4 \pi D_{L}^{2}}{(1+z)^{2+\alpha}}
\label{eq:energy_r147}
\end{equation}
where $\mathcal{F}$ is the fluence (see Table~\ref{tab:hyperflash_burst_prop_appendix}), $D_{L}$ is the luminosity distance, $z$ is the redshift, and $\alpha$ is the spectral index. The factor $(1+z)^{2+\alpha}$ accounts for the redshift correction. We adopt $\alpha = 0$ which is consistent with the assumption used in the fluence calculation ($\mathcal{F}_{\nu} \propto \nu^{\alpha}$). For \ronefourseven, we use $z = 0.130287$, corresponding to a luminosity distance of $D_{L} = 616$\,Mpc \citep{bhardwaj_2025_apjl}.

In addition to the burst energies reported here, we incorporate fluences from observations with FAST \citep{zhang_2025_arxiv}, which detected $11{,}553$ bursts over $33.86$\,hours between 2024 January 28 (MJD~$60337$) and 2024 August 29 (MJD~$60552$). We further include $673$ NRT bursts from J.~Huang~in prep. Since both observational campaigns were conducted at L-band ($\sim$$1.4$\,GHz) during overlapping time periods, this enables a direct comparison with our reported results, and avoids potential differences due to a frequency- or time-dependent burst energy distribution.

To determine the lower bound above which the cumulative distribution is well described by a single power-law, we use the Python package \texttt{powerlaw} \citep{alstott_2014_ploso}. We find turnover points of $3.1$\,Jy\,ms ($1.1\times10^{30}$\,\erghz) for FAST and $1.1$\,Jy\,ms ($3.8\times10^{29}$\,\erghz) for NRT. For HyperFlash, we combine all bursts detected with Westerbork, Onsala, \torun, and Stockert, and in cases of co-detections by multiple telescopes, we retain only the brightest detection. We adopt the completeness threshold of the least sensitive instrument (Westerbork) at $24.4$\,Jy\,ms as a lower limit, and identify a turnover at $25.2$\,Jy\,ms ($8.9\times10^{30}$\,\erghz). Following this, we estimate an initial value for the power-law slope $\gamma$ using a maximum likelihood approach \citep{crawford_1970_apj,james_2019_mnras}. We then fit the power-law using \texttt{scipy.optimize.curvefit}, assuming a conservative $20$\,\% uncertainty on the SEFD. This uncertainty is consistent with values adopted in studies using similar recording setups \citep[e.g.][]{kirsten_2021_natas, ouldboukattine_2026_mnras_2}. We apply the same uncertainty to the FAST and NRT burst samples to ensure a consistent comparison between the different datasets. For the uncertainty of $\gamma$, we quote two errors: the $1\sigma$ error reported by \texttt{curvefit}, and an error obtained by a resampling method and drawing $1000$ random subsets of $90$\,\% of the bursts with replacement and refitting the power-law each time.

\begin{figure*}
    \centering
    \includegraphics[width=\linewidth]{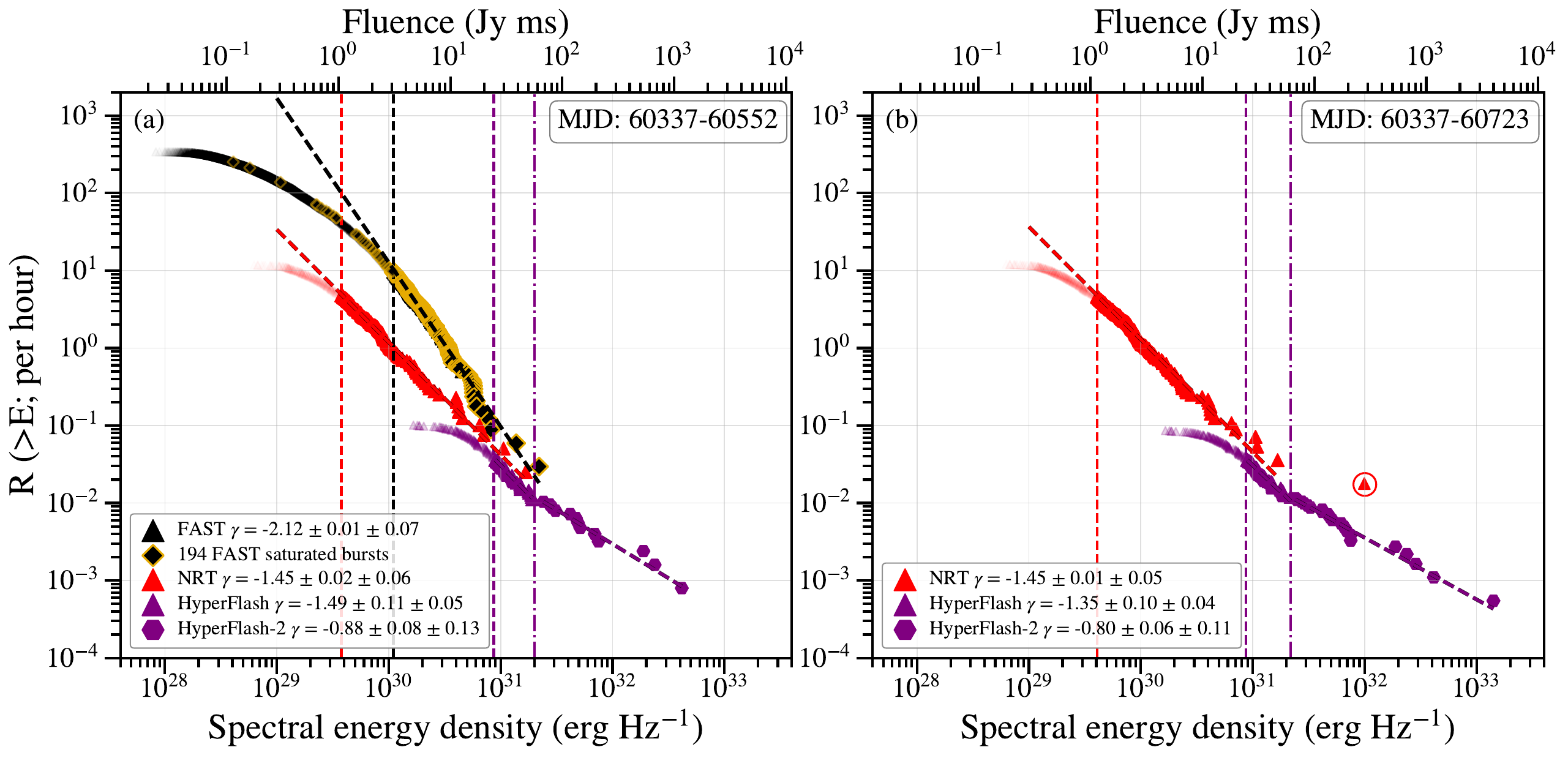}
    \caption{Cumulative burst rate distribution of isotropic-equivalent spectral energy densities for bursts detected at L-band ($\sim$$1.4$\,GHz). HyperFlash bursts (purple) consist of bursts detected with Onsala, \torun, Stockert and Westerbork. The black data points are bursts detected with FAST and the red represent bursts detected by NRT. We constrain the burst window to the FAST burst-detection period between MJD~$60337-60551$ in the left panel (a), and to the NRT campaign period between MJD~$60337-60723$ in the right panel (b). This allows for a fairer comparison between detections of different instruments, in case the energy distribution is frequency- or time-dependent. The red, black and purple vertical lines represent the thresholds above which the respective energy distributions are best described by a single power-law, as determined by the Python package \texttt{powerlaw}. The purple dash-dotted vertical line denotes the break in the energy distribution of HyperFlash at $2.4\times10^{31}$\,erg\,Hz$^{-1}$. Transparent data points to the left of the vertical dashed lines are excluded from the fit. We assume a $20$\,\% error on burst energies. The first error represents the $1\sigma$ statistical uncertainty on the fit and the second error is the $1\sigma$ error after the bootstrapping method (see main text). The most energetic burst in our sample, ``The STROOP'', is the rightmost data point in the right panel.}
    \label{fig:burst_distr_r147}
\end{figure*}

In Figure~\ref{fig:burst_distr_r147}, we show the fitted power-laws to the cumulative energy distributions for FAST, NRT, and HyperFlash. In the left panel (a), we restrict the distributions to the time span of the FAST campaign, covering $216$~days between MJD $60337-60552$, and include only bursts detected within this interval. For FAST, this corresponds to the full sample of $11{,}553$ bursts, whereas NRT and HyperFlash contribute $468$ and $127$ bursts, respectively. The $194$~bursts identified as saturated by \cite{zhang_2025_arxiv} are shown as diamonds with a golden border. For FAST, we find $\gamma^{\mathrm{FAST}} = -2.12 \pm 0.01 \pm 0.07$, and for \nancay $\gamma^{\mathrm{NRT}} = -1.45 \pm 0.02 \pm 0.06$. For HyperFlash, we observe a clear break in the energy distribution and therefore fit a broken power-law. The break occurs at $2.4\times10^{31}$\,\erghz ($68$\,Jy\,ms). We then fit two power-laws using the same methodology, obtaining $\gamma^{\mathrm{HF-1}} = -1.49 \pm 0.11 \pm 0.05$ below the break and $\gamma^{\mathrm{HF-2}} = -0.88 \pm 0.08 \pm 0.13$ above it.

In the right panel (b), we instead consider the full time span of the \eclat campaign, spanning $387$~days between MJD $60344-60723$. We now make use of the full sample of $673$~\eclat detected bursts, with HyperFlash contributing $157$ bursts. For NRT, we find $\gamma^{\mathrm{NRT}} = -1.45 \pm 0.01 \pm 0.05$, excluding the final data point highlighted with a red circle. This point deviates from the distribution; however, due to the insufficient number of data points, we do not fit a broken power-law. This deviation could be explained by the burst originating from an underlying broken power-law distribution. For HyperFlash we do fit a broken power-law and find $\gamma^{\mathrm{HF-1}} = -1.35 \pm 0.10 \pm 0.04$ below the break and $\gamma^{\mathrm{HF-2}} = -0.80 \pm 0.06 \pm 0.12$ above it.

\subsection{Variable dispersion measure} \label{sec:variable_dm}

\begin{figure}
    \centering
    \includegraphics[width=\linewidth]{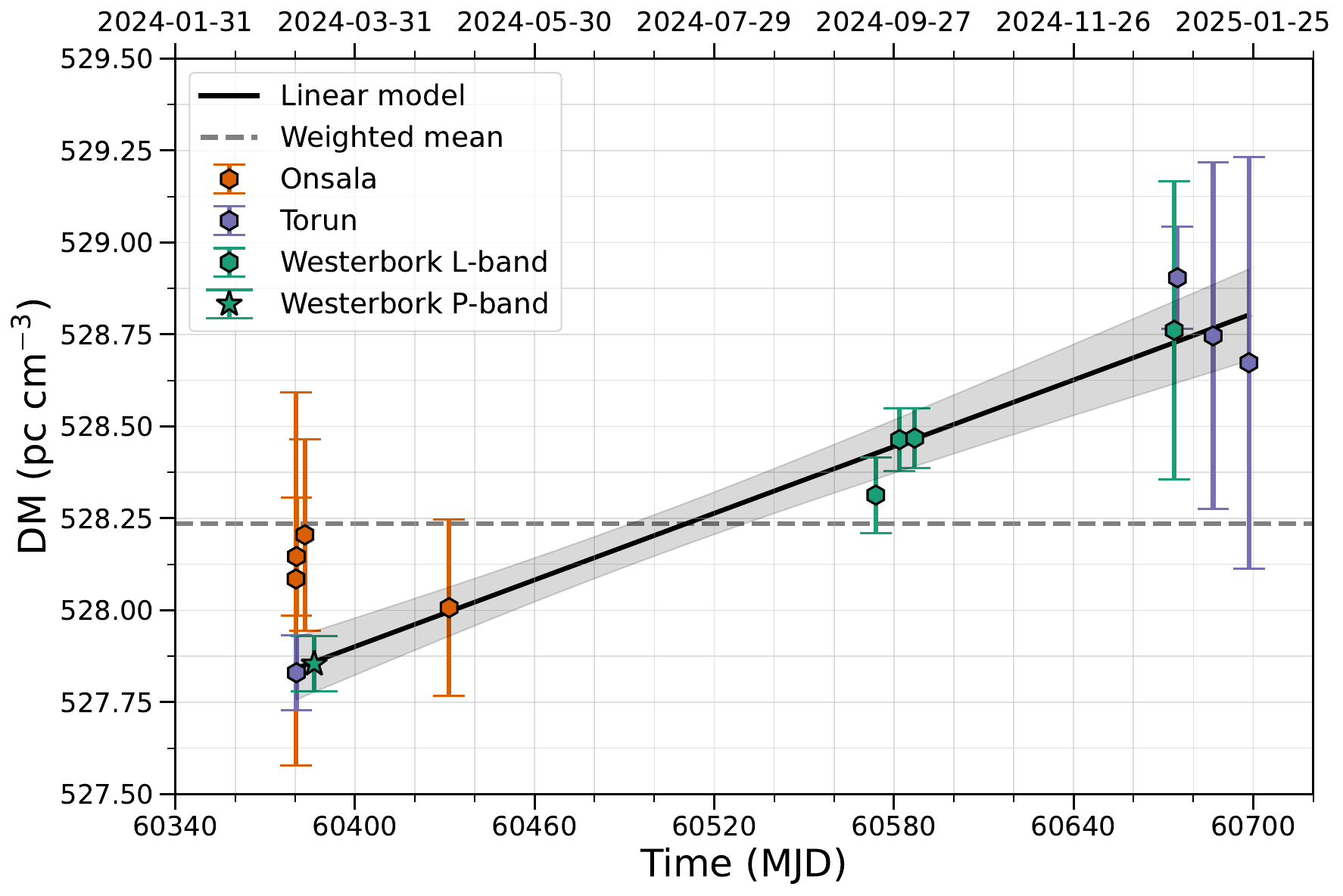}
    \caption{DM evolution of $13$~bursts from \ronefourseven observed at L-band ($\sim$$1.4$\,GHz; $12$ bursts) and P-band ($324$\,MHz; 1 burst) detected with Onsala, Westerbork, and \torun. The bursts were selected based on the presence of temporal microstructure (emission on timescales $<100$\,\microsec) and the availability of baseband data, which enabled precise DM measurements (see Appendix Figure~\ref{fig:r147_dynamic_matrix} for a selection of the dynamic spectra and Appendix Table~\ref{tab:evolving_dm} for the values). The DM of \ronefourseven increased linearly by $+0.96 \pm 0.06$\,\dmunit over a period of $318$\,days (see Section~\ref{sec:variable_dm}).}
    \label{fig:dm_evolution}
\end{figure}

The high S/N of a subset of the bursts, in combination with the availability of raw voltage data, allow us to investigate DM variations between bursts. Westerbork, \torun, and Onsala record raw voltage data. These data, in combination with the Super FX Correlator (SFXC) software \citep{keimpema_2015_exa}, allow us to create \texttt{filterbank} files with time resolutions as high as $31.25$\,ns, while also applying coherent dedispersion; i.e., correcting for dispersion within individual frequency channels. 

We find that $13$~bursts in our sample with available baseband data show unresolved structure in the time series on timescales of $10$–$100$\,\microsec. Our sample consists of four bursts detected with Onsala, five with Westerbork (including one at P-band), and four with \torun. As described previously in \citet{ouldboukattine_2026_MNRAS}, we optimize the DM by fitting a Gaussian profile to the S/N–DM trial curve, focussing on narrow and bright components. This assumes that the highest S/N corresponds to the optimal DM, since the burst component is mostly unresolved; visual inspection confirms that this works well. From the voltage data, we create coherently dedispersed \texttt{filterbank} files with SFXC for a range of trial DM values, sampled in steps of $0.01-0.1$\,\dmunit depending on the width of the component. For each trial DM, we measure the peak S/N of the component. We then fit a Gaussian profile to this curve to determine the optimal DM and its uncertainty. We define the DM uncertainty as the DM interval over which the fitted S/N–DM Gaussian decreases by $1$ in S/N from its maximum value.
In Figure~\ref{fig:dm_evolution}, we show the measured DM values and their uncertainties as a function of time. In Appendix Figure~\ref{fig:r147_dynamic_matrix}, we show the dynamic spectra dedispersed at the best-fit DM, as well as at the best-fit DM $\pm 3\sigma$\,\dmunit for the five brightest bursts. We also show the S/N vs. DM trial curves with the Gaussian fits, plotted on a shared x-axis to highlight the measured evolution of DM over time. The DM of \ronefourseven increased by $+0.96 \pm 0.06$\,\dmunit over $318$\,days between MJD $60380-60698$.

For the $13$~bursts in our sample for which we fit DMs, and given those bursts' arrival times and their estimated DM uncertainties, we explore this observed variability. We first find the maximum likelihood parameters for a constant DM model with time (the weighted mean) and a linear model of DM versus time, and compute the fits' associated Bayesian Information Criterion (BIC), which is a measure of the goodness of fit that penalizes for the complexity of the model. The linear model (shown in Figure~\ref{fig:dm_evolution}) is strongly preferred to the constant model for our DM versus burst arrival time data, with BIC $-6.0$ and $68.7$ for the linear and constant best-fit model, respectively.  
While this can be considered strong evidence for a variable DM model, we further perform a simulation to diagnose if the observed DM measurements could align in a way that is as well described by a linear model by chance, i.e., could the measurements reasonably approximate linearity to this degree due to stochastic variability beyond that which is captured by the estimated DM uncertainties. Following \cite{cook_2026_arxiv}, we fix the observed burst arrival times and estimated DM uncertainties from our high-S/N bursts with sharp features, and then draw $13$ simulated DMs from a normal distribution with the same mean and variance as the subsample. For each simulated sample, we compute the BIC for the best-fit linear model, and then record how many times that BIC suggests a fit as good or better than the fit derived for the observed data. We find that the simulation produces random draws that are as well described by a linear model only $0.27$\% of the time ($2.7$\,$\sigma$ Gaussian equivalent). We thus conclude that the observed DM variation is unlikely to arise from stochastic variation alone and instead that the linear trend we see is statistically significant. 

\section{Discussion}
Here we discuss \ronefourseven's energetics, focusing on burst rate vs. energy and the observed cumulative energy release at radio wavelengths over $806$~days. We also discuss the source's time-variable DM as a means of constraining its local environment. We conclude the discussion comparing to the energetics of magnetar bursts observed in X-rays and $\gamma$-rays.

\subsection{A break in the cumulative energy distribution}
We identify a break in the cumulative energy distribution of \ronefourseven at $2.4\times10^{31}$\,\erghz ($68$\,Jy\,ms; Figure~\ref{fig:burst_distr_r147}). This break is obtained by fitting a broken power-law to the HyperFlash sample above the completeness threshold (Section~\ref{sec:burstrate_how}), yielding slopes of $\gamma_\textrm{HF-1}=-1.49\pm0.11\pm0.05$ below the break and $\gamma_\textrm{HF-2}=-0.88\pm0.08\pm0.13$ above it. The lower-energy slope of HyperFlash reflects the steep part of the burst distributions from NRT ($\gamma_\textrm{NRT}=-1.45\pm0.02\pm0.06$) and FAST ($\gamma_\textrm{FAST}=-2.12\pm0.01\pm0.07$). 

In the right panel in Figure~\ref{fig:burst_distr_r147}, we also show the cumulative distribution for bursts detected during the full NRT campaign, which spans an additional $171$\,days and adds $205$ and $30$ bursts to the NRT and HyperFlash samples, respectively. This increased exposure reveals an additional bright NRT burst and a bright detection we refer to as “the STROOP” by HyperFlash (Figure~\ref{fig:the_stroop}; Section~\ref{sec:disc_energy_stroop}). The bright NRT burst lies above the fitted distribution and appears to be consistent with the break observed in the HyperFlash sample. However, given the limited statistics at the high-energy end of the NRT distribution due to lower on-source time, we do not attempt a broken power-law fit to the NRT data.

The observed break is consistent with that reported by \cite{huang_2025_raa}, who find a break at $\sim$$2\times10^{31}$\,\erghz after 318\,hours of high-cadence observations with the Kunming $40$-meter Radio Telescope (KM40M) between 2024 March 8 (MJD~$60377$) and 2024 November 5 (MJD~$60619$) at S-band ($2.2$\,GHz). \cite{wang_2025_apj} report a slope of $\gamma_\textrm{TMRT} = -1.20 \pm 0.03 \pm 0.02$ at $7.5\times10^{37}$\,erg (corresponding to $\sim$$7.5\times10^{29}$\,\erghz), based on $155$ bursts detected above a fluence threshold of $0.72$\,Jy\,ms over $182$\,hours of observations at S-band ($2.25$\,GHz) with the Shanghai Tianma Radio Telescope (TMRT) between 2024 January 29 (MJD~$60338$) and 2025 February 15 (MJD~$60721$). The resulting power-law index of the TMRT is slightly flatter than both the $\gamma_\textrm{HF-1}$ and $\gamma_\textrm{NRT}$ slopes shown in panel~b of Figure~\ref{fig:burst_distr_r147}. The NRT S-band burst sample may help address whether the power-law index could be frequency-dependent and is consistent with the S-band TMRT results (J.~Huang~in prep.).

The NRT and FAST power-law slopes are inconsistent (Figure~\ref{fig:burst_distr_r147}). The NRT slope is, however, consistent with the HyperFlash results for both of the time ranges we consider. The discrepancy between NRT and FAST may arise from non-overlapping observing windows combined with the relatively short duration of both campaigns during periods of highly variable burst activity from \ronefourseven \citep{zhang_2025_arxiv}. In particular, NRT missed a $\sim$$10$-day active phase around MJD~$60375$, during which the source was significantly more active (see inset in Figure~\ref{fig:cumulative_sum_energy_hf_nrt_fast}). Additionally, \cite{zhang_2025_arxiv} report that $194$ bursts in the FAST sample were affected by saturation due to their high brightness; we have highlighted these bursts with a golden border in the cumulative distribution in the left panel of Figure~\ref{fig:burst_distr_r147}. As also noted by \cite{huang_2025_raa}, saturation of these events may lead to an underestimation of their burst energies, which would artificially steepen the observed distribution. The actual cumulative distribution of FAST may therefore be flatter, and more consistent with the NRT power-law.

High-cadence observations of highly active FRB sources have revealed breaks in the energy distributions of four repeaters. We summarise the corresponding power-law indices in Table~\ref{tab:repeater-breaks}. For two repeaters, \rsixseven and \rmkt, the break is only evident when comparing slopes obtained with multiple telescopes \citep{kirsten_2024_natas, ouldboukattine_2026_MNRAS}. For the other two sources, \roneoneseven and \ronefourseven (this work), we identify a break using single-telescope data, which is further supported by comparing to data from other observing campaigns \citep{ouldboukattine_2026_mnras_2}. The ability to detect such a break depends critically on sufficient on-source observing time (hundreds to thousands of hours) to detect rare high-energy bursts. These observed breaks in the burst energy distributions may reflect multiple emission processes contributing to burst production, where the rate of high-energy bursts is higher than would be expected by extrapolating based on the distribution of lower-energy bursts. We discuss this further in the next section.

\begin{table*}
\caption{Breaks in the cumulative energy distributions (in specific energy and fluence) and the corresponding power-law slopes below and above each break, for four hyperactive repeaters.}
\label{tab:repeater-breaks}
\resizebox{\textwidth}{!}{%
\begin{tabular}{l c c c c l}
\hline
\hline
Source                        & $E_{\nu, \textrm{break}}$ [erg/Hz]                         & $\mathcal{F}$-break [Jy ms] & $\gamma_\textrm{low} < E_{\nu, \textrm{break}}$                          & $\gamma_\textrm{high} > E_{\nu, \textrm{break}}$                                            & Reference                     \\ \hline
FRB~20201124A$\mathrm{^{a}}$     & $5.0-8.0~\times~10^{30}$     & $24 - 39$           & $\gamma_\textrm{FAST} = -1.94 \pm 0.10 \pm 0.06$  & $\gamma_\textrm{HF} = -0.48 \pm 0.11 \pm 0.03$ & \cite{kirsten_2024_natas}            \\
FRB~20220912A$\mathrm{^{b}}$  & $3.2~\times~10^{30}$         & $24$                & $\gamma_\textrm{NRT} = -1.69 \pm 0.01 \pm 0.06$  & $\gamma_\textrm{NRT} = -0.60 \pm 0.08 \pm 0.05$ & \cite{ouldboukattine_2026_mnras_2}   \\
FRB~20240619D$\mathrm{^{a}}$          & $-$$\mathrm{^{c}}$                            & $\sim 25$           & $\gamma_\textrm{NRT} = -1.87 \pm 0.03 \pm 0.09$  & $\gamma_\textrm{HF} = -0.77 \pm 0.11 \pm 0.14$ & \cite{ouldboukattine_2026_MNRAS}     \\ \hline
FRB~20240114A$\mathrm{^{b}}$ & $2.4~\times~10^{31}$     & $68$                & $\gamma_\textrm{HF} = -1.35 \pm 0.10 \pm 0.04$  & $\gamma_\textrm{HF} = -0.88 \pm 0.06 \pm 0.12$ & This work \\ \hline
\multicolumn{6}{l}{$\mathrm{^{a}}$$E_{\nu, \textrm{break}}$ inferred from comparing power-law slopes across multiple telescopes.} \\
\multicolumn{6}{l}{$\mathrm{^{b}}$$E_{\nu, \textrm{break}}$ obtained from fitting a broken power-law to a single distribution.} \\
\multicolumn{6}{l}{$\mathrm{^{c}}$At the time of writing, no redshift is known for \rmkt, which prevents conversion of fluence into energy.} \\
\end{tabular}%
}
\end{table*}

\subsection{Interpreting the observed break in the energy distribution}
\label{sect:interpretation_break}
A break in the cumulative energy distribution at $E_{\nu, \textrm{break}}\sim2.4\times10^{31}$\,\erghz (see Figure~\ref{fig:burst_distr_r147} and Table~\ref{tab:repeater-breaks}) now appears to be ubiquitous amongst hyperactive repeating FRBs. However, the origin of this break is unclear. It may be the direct result of fundamentally different mechanisms acting at the highest energies, reflecting a change in the intrinsic, energy-dependent burst rate. On the other hand, it could be the result of structure in the FRB beaming distribution --- since all the energies we quote in this paper assume isotropic emission. In this case, the true burst rate distribution may follow a single power-law, but a break arises when this is convolved with a beaming distribution. 
\par
Comparisons are often made between repeating FRBs, magnetar X-ray bursts, and terrestrial earthquakes --- especially with respect to wait-time distributions and cumulative energy distributions (\citealt{cheng_1996,wang_2017,wang_2018_starquake,totani_2023}; see Section \ref{sect:magnetars}). However, no high-energy flattening of the terrestrial earthquake cumulative energy distribution has been observed, and this is not generally predicted within models of critical systems proposed for magnetar and earthquake behaviour \citep{bak_1987,aschwanden_2016}. In this subsection, we investigate whether the observed burst rate energy distribution of hyperactive repeating FRBs could be explained by beaming effects alone.
\par
The true (intrinsic) energy of a radio burst is:
\begin{equation}
    E_{\rm true} = f_{\rm b} E_{\rm iso} = f_{\rm b} E_{\rm obs}, 
\end{equation}

where $f_{\rm b} = \Omega_{\rm b}/4 \pi$ and $\Omega_{\rm b}$ is the beaming solid angle. The $E_{\rm obs}$ and $E_{\rm iso}$ are equivalent: the derived energy assuming isotropic emission is observed. We assume in the following that FRBs are emitted in random directions from the source, plausibly stemming from a variety of emission locations with respect to the central source. This is motivated in part by the non-detection of a spin period from even the most active repeaters. This assumption means that the probability of observing a particular FRB is equivalent to the beaming factor $f_{\rm b}$: $P_{\rm obs} = f_{\rm b}$. This means that for $f_{\rm b} = 1$ the burst is isotropic and always observed. We specify a minimum beaming fraction $0 <f_{\rm b, min} < f_{\rm b}$ to ensure normalizability.
\par
We consider four cases, corresponding to different assumptions for the FRB energy distribution $\Phi(E_{\rm true})$ and the beaming distribution $p(f_{\rm b})$. In Appendix~\ref{app:interpretation_break_maths}, we calculate the observed burst rate distribution $d\dot{N}/dE_{\rm obs}$ for each of these cases, the main results of which are summarised in Table \ref{tab:theorybreaks}. We find that to recover the broken power-law in the observed cumulative burst rate distribution, a break in the true energy rate distribution of emitted FRBs is required. In other words, a break in the beaming probability distribution cannot explain the observed high-energy break in the energy rate distribution in Figure~\ref{fig:burst_distr_r147}. This indicates that there may be two distinct mechanisms powering FRBs from active repeaters at low and high burst energies, with a transition around $E_{\nu, \textrm{break}}\sim10^{30-31}$\,\erghz. Further work is required to fully explore the effect of beaming on the observed burst distributions, including consideration of additional functional forms of $\Phi(E_{\rm true})$ and $p(f_{\rm b})$, scenarios in which these distributions are not independent, as well as the angular structure of individual FRB beams. 

\begin{table}
\centering
\caption{Summary of resulting observed cumulative burst rate for four cases varying true burst energy and beaming distributions. Only a broken power-law in the burst energy distribution $\Phi(E_{\rm true})$ results in an observed broken power-law as found for hyperactive repeaters. Full details can be found in Appendix \ref{app:interpretation_break_maths}.}
\label{tab:theorybreaks}
\resizebox{\columnwidth}{!}{%
\begin{tabular}{cccc}
\toprule
Case & $\Phi(E_{\textrm{true}})$ & $p(f_{\textrm{b}})$ & Observed $d\dot{N}/dE_{\textrm{obs}}$ \\
\midrule

1 & Fixed & PL $\big(f_{\textrm{b}}^{-\alpha}\big)$ & PL $\big(E_{\textrm{obs}}^{-(3-\alpha)}\big)$ \\

2 & PL $\big(E_{\textrm{true}}^{-\gamma}\big)$ & PL $\big(f_{\textrm{b}}^{-\alpha}\big)$ & PL $\big(E_{\textrm{obs}}^{-\gamma}\big)$ \\

3 & PL $\big(E_{\textrm{true}}^{-\gamma}\big)$ &
\makecell[c]{BPL -- App.~Eq.~\ref{app_eq:case3_bpl}} &
PL $\big(E_{\textrm{obs}}^{-\gamma}\big)$ \\

4 &
\makecell[c]{BPL -- App.~Eq.~\ref{app_eq:case4_bpl_true}} &
PL $\big(f_{\textrm{b}}^{-\alpha}\big)$ &
\makecell[c]{BPL + Transition -- App.~Eq.~\ref{app_eq:case4_bpl_observed}} \\

\bottomrule
\multicolumn{4}{l}{$\mathrm{^{*}}$Powerlaw (PL) \& Broken powerlaw (BPL)} \\
\end{tabular}%
}

\end{table}

\subsection{Energy release as a function of time} \label{sec:disc_energy_stroop}
The cumulative sum of isotropic energies for bursts detected at L-band ($\sim$$1.4$\,GHz), shown in Figure~\ref{fig:cumulative_sum_energy_hf_nrt_fast}, implies that \ronefourseven released a large amount of energy during its different periods of burst activity. At a measured redshift of $z=0.130287$, and after scaling to the canonical assumed bandwidth of $1$\,GHz, we infer the total emitted radio energy. The FAST burst sample, consisting of $11{,}553$ bursts, corresponds to a total cumulative energy of $E^{\mathrm{FAST}}_{\mathrm{tot}} = 2.3 \times 10^{42}$\,erg \citep{zhang_2025_arxiv}. Remarkably, despite consisting of only $130$~bursts, the HyperFlash sample {\it observed over the same span as FAST} reaches a cumulative energy of $E^{\mathrm{HF}} = 2.1 \times 10^{42}$\,erg --- comparable to the FAST sample within the same observing time window (see Figure~\ref{fig:cumulative_sum_energy_hf_nrt_fast}). This indicates that the brightest bursts ($\gtrsim 25$\,Jy\,ms), which are under-represented in FAST observations due to their rarity, contribute in roughly equal measure to the sum of cumulative burst energies. This behaviour is reminiscent of magnetar bursts detected in X-rays and $\gamma$-rays, in which the energetic output of a substantial burst storm of thousands of short X-ray bursts can be eclipsed by a single intermediate or giant flare (see Section \ref{sect:magnetars} for an elaboration of this discussion). 

During {\it all observed} periods of activity, spanning $478$~days (see Appendix~Figure~\ref{fig:obs_campaign_r147}), we were observing \ronefourseven for $17$\,\% of the time, an exceptionally high on-source exposure, and detected $178$ unique bursts. The cumulative isotropic-equivalent energy of the whole 178-burst HyperFlash sample is $E^{\mathrm{HF}}_{\mathrm{tot}} = 4.4 \times 10^{42}$\,erg. Following several active periods, the source was not detected in the last 328 days of our observing campaign, despite continued regular observations (see Appendix~Figure~\ref{fig:cumulative_sum_energy_hf}). This intermittent detectability, even after phases in which the source significantly contributes to the all-sky FRB rate as a whole, is consistent with behaviour observed in other repeating FRBs (e.g., \citealt{ouldboukattine_2026_MNRAS}).

Following the detection of the STROOP, we did not detect any subsequent bursts for $68$~days, despite regular, near-daily observations of $3-4$\,hours each (see Appendix Figure~\ref{fig:cumulative_sum_energy_hf}). However, we do not have sufficient observational coverage before and after the detection of the STROOP to constrain whether it halted the source’s bursting activity, for example due to the large instantaneous depletion of the energy reservoir, and we can therefore only speculate that the lower burst rate after the STROOP is a consequence of this event. Nonetheless, we note that similar behaviour was observed following the giant magnetar flare of SGR~1806$-$20 in December 2004 \citep{palmer_2005}, which marked the end of high burst activity \citep{gotz_2006}. However, we note that this is in direct contrast to the giant flare from SGR~1900+14 \citep{hurley_1999}, in which the giant flare marked the onset of renewed burst activity \citep{imbrahim_2001}.

\subsubsection*{The STROOP}
The STROOP (Figure~\ref{fig:the_stroop}) is arguably the most energetic FRB detected to date, including apparently one-off events, with an isotropic-equivalent energy of $E_{\rm iso} = 1.4 \times 10^{42}$\,erg. It accounts for $32\%$ of the total cumulative energy in the HyperFlash \ronefourseven burst sample, and is an order of magnitude higher energy than even the most distant FRB discovered at $z=2.15$ \citep{caleb_2025_arxiv}. Nonetheless, its energy is consistent with the apparent limit in maximum burst energy found in studies of both repeating and apparently one-off FRBs; see \citealt{ouldboukattine_2026_mnras_2}, and references therein.
Additionally, based on thousands of FRBs from the CHIME/FRB Catalogue 2 \citep{chime_2026_cat2}, \cite{shah_2026} derive a lower limit on the maximum isotropic FRB energy of $1.2^{+0.7}_{-1.1} \times 10^{42} \, {\rm erg}$, which is consistent with the energy of the STROOP. They further show that many hundreds of bursts cluster around the inferred energy of $E_{\rm iso} \sim 10^{42} \, {\rm erg}$, independent of DM-implied redshifts, suggesting a possible physical upper limit at an energy comparable to the STROOP.

In addition to total burst energy, peak luminosity is another physically constraining observable. We calculate the peak luminosity of the STROOP by converting the peak S/N value of the brightest component where it is just barely resolved temporally, at $16$\,\microsec, and find $1236$\,Jy. We convert this peak flux ($S_\nu$) to luminosity following Equation~$7$ from \cite{macquart_2018_mnras};
\begin{equation}
    L_\nu = \frac{4 \pi D_\mathrm{L}^2}{1+z} \times S_\nu\,\,\,\mathrm{erg\,s^{-1}\, Hz^{-1}}
\end{equation}
and find $4.9\times10^{35}$\,\erghzs. Following Equation~$8$ from \cite{zhang_2018_apjl} we next convert this value to a peak luminosity,
\begin{equation}
    L_\textrm{peak, iso} = L_\nu \times \nu_\textrm{c}\,\,\,\mathrm{erg\,s^{-1}}
\end{equation}
Here $\nu_\textrm{c}$ is the central frequency, which for the STROOP is $1.271$\,GHz, and find $L_{\rm peak, iso} = 6.4 \times 10^{44} \, {\rm erg \, s^{-1}}$. 
\par
Some proposed FRB emission mechanisms have derived maximum energies/luminosities corresponding to either maximal trigger events, depletion of available energy, or microphysical constraints. \cite{lu_maximum_2019} set a maximum FRB luminosity within a coherent curvature radiation framework, due to the rapid screening of the parallel electric field by Schwinger pairs above $E_{\parallel} \sim 2.5 \times 10^{12} \, {\rm esu}$: 
\begin{equation}
\label{eq:lu_constraints}
  L_{\rm max} \sim 2 \times 10^{47} \, {\rm erg\, s^{-1}} \; {\rm min} \big(\rho_6^{2}, B_{16}\, \rho_6^{4/3}\, \nu_{\rm GHz}^{-2/3} \, \big),
\end{equation}
where $\rho$ is the curvature radius of magnetic field lines along which particles accelerate (in cm), $B$ is the local magnetic field strength (in Gauss), and $\nu$ is the observing frequency (in GHz). Within the framework of this specific radiation model, the STROOP requires a minimum magnetic field of $B \gtrsim 3.5\times 10^{13} \, {\rm G \, \rho_6^{2}}$, suggestive of a magnetar progenitor. We show this limit in Appendix Figure~\ref{fig:ccr_magnetospheric_limits}, alongside similar period-dependent constraints specific to coherent curvature radiation from Equation~15 in \citet{cooper_2021_mnras}. These simplified constraints are complicated by the unknown radio efficiency, $\eta_{\rm r}$, which may increase magnetic field requirements, but also unknown propagation effects that may artificially increase instantaneous luminosities within sub-burst structures, decreasing field requirements \citep{sobacchi_2024}. More generically, many magnetar models of FRB emission predict counterpart emission at other wavelengths, scaling linearly with the total flare energy. This includes maser shock models \citep{metzger_2019_mnras}, in which $E_{\rm FRB, peak} \sim  10^{42} \, {\rm  erg \,  s^{-1}} f_{\xi, -3} \, E_{\rm flare, 45}$, where $f_{\xi, -3}$ is the radiative efficiency.

\subsection{DM increasing in time and constraining the magnetic field strength}
Using the fine temporal structure in a subsample of $13$~bursts for which baseband data were saved, we probe \ronefourseven's changing DM (see Section~\ref{sec:variable_dm}). We find that the DM is linearly increasing by $+0.96 \pm 0.06$\,\dmunit over a period of $318$\,days between MJD~$60380$ and $60698$; see Figure~\ref{fig:dm_evolution} and Appendix Figure~\ref{fig:r147_dynamic_matrix}. The magnitude of this increase falls within the range of DM uncertainties of other observing campaigns targetting \ronefourseven --- e.g., FAST: $529.1\pm1.3$\,\dmunit \citep{zhang_2025_arxiv} and uGMRT: $528.36^{+1.88}_{-1.11}$\,\dmunit \citep{panda_2025_apj} --- and hence we report what we consider to be the first robust measurement of \ronefourseven's time-variable DM. Previously, we took the same approach in measuring the change in DM for \rmkt, where we found an increase of $+0.41 \pm 0.016$\,\dmunit over a period of two months \citep{ouldboukattine_2026_MNRAS}. For \ronefourseven, detecting a comparable increase on similar timescales over 2 months, would have been very challenging, as the microstructure in its bursts is broader than that of bursts from \rmkt. For \rmkt, the temporal structure can be as narrow as $1$–$8$\,\microsec, whereas for \ronefourseven the finest structure we observe is on the $8$–$64$\,\microsec scale. These wider temporal scales lead to larger uncertainties in the DM measurements. However, over longer timescales, the DM increase for \ronefourseven becomes measurable due to its extended period of activity.

Long-term monitoring (months to years), combined with sustained activity from repeating FRB sources and the presence of microstructure in burst time profiles, is essential for constraining DM variations. To date, DM variations and evolution have only been observed in a handful of repeating sources, including \rone \citep{snelders_2025_arxiv}, \roneoneseven \citep{abbott_2026_arxiv}, \ronezeronine \citep{pandhi_2026_apjl}, and \rmkt \citep{ouldboukattine_2026_MNRAS}. Not all repeating sources have observed activity over a sufficiently long baseline. Some repeating sources from CHIME/FRB's second catalogue have exhibited measurable DM variability over their only weeks- to months-long active periods, as illustrated in Figure~5 of \cite{cook_2026_arxiv}. Therefore, for most repeaters, it is challenging to constrain possible DM variations, especially when $\Delta\textrm{DM}<1$\,\dmunityr. 

A consistently active repeating source is required to identify trends in DM. This is illustrated in Figure~5 of \cite{snelders_2025_arxiv}, which presents the DM evolution of \rone spanning more than 10 years. In contrast, \rthree has remained consistently detectable, with its DM remaining stable within $\lesssim 0.8$\,\dmunit over a baseline of three years \citep[$2019-2022$;][]{sand_2023_apj,mckinven_2023_apj}. Smaller DM variations have been reported (e.g., FRB~20190907A\xspace: $\sim +0.15$\,\dmunityr; \citealt{cook_2026_arxiv}), which could suggest that the DM of \rthree is evolving, but that we currently lack the time span and sufficiently narrow bursts to constrain this. Notably, \rthree produces bursts with unresolved temporal structure down to $1$\,\microsec \citep{nimmo_2021_natas}. Continued monitoring and the detection of more bursts with such fine temporal features from \rthree could help determine whether the DM is truly stochastically stable or evolving over time. 
The limited detectability of repeating sources over longer baselines with sufficient S/N and resolved microstructure is insufficient to establish whether DM evolution is a universal property of repeating FRBs, and warrants continued observational follow-up, necessarily with high-time-resolution data.

FAST and Effelsberg observations of \ronefourseven by \citet{wang_2026_apjs} and \citet{manaswini_2026_arxiv}, respectively, indicate day-to-day RM fluctuations that are coincident with the DM variations presented in this work. Between MJD $60380$--$60700$, during which we measure a DM increase of $0.96 \pm 0.06$\,\dmunit, the RM varies stochastically between $\sim350$–$400$\,\rmunit. Making the simplifying assumption that the same intervening plasma components are responsible for both RM and DM variations (see, e.g., \citealt{michilli_2018_natur, ouldboukattine_2026_MNRAS}), we can approximate the spatially averaged parallel magnetic field component as:

\begin{equation}
B_{\parallel,{\rm \upmu G}} = 1.23\,\upmu\textrm{G}\,\left( \frac{\Delta\textrm{RM}}{\textrm{rad\,m}^{-2}}\right) \left( \frac{\Delta\textrm{DM}}{\textrm{pc\,cm$^{-3}$}}\right)^{-1}\sim 64\, {\rm \upmu G}
\end{equation}

The inferred $B_{\parallel}$ is lower than other actively repeating FRBs (e.g., \citealt{michilli_2018_natur,annathomas_2023_sci,moroianu_2026_apj,ouldboukattine_2026_MNRAS,pandhi_2026_apjl}), indicating a range of magnetic environments may be present for active repeaters. We caution that, in this case and \textit{during this specific time range}, the RM variations are not monotonic, in contrast to previous work \citep{ouldboukattine_2026_MNRAS}. This hinders a more robust estimate of $B_{\parallel}$, but may indicate a turbulent, albeit less extreme, environment compared to highly active repeaters. \citet{wang_2026_apjs} further report a decrease in RM of $\sim130$\,\rmunit after MJD~$60700$ in the FAST burst sample. Although we detect bursts in this time range between MJD~$60700$–$60820$, no bursts exhibit unresolved temporal features. This prevents us from further constraining the DM evolution during the final $120$\,days of activity, and consequently from approximating the parallel magnetic field strength during this period of RM variation.

\subsection{An analogue to magnetar flares} \label{sect:magnetars}
Magnetars are the strongest candidates for the sources of (repeating) FRBs and a connection between high-energy magnetar bursts and FRBs has been conjectured by many authors (e.g., \citealt{popov_2010,lyubarsky_2014_mnras,2019ApJ...879....4W,bransgrove_2026}). Broadly speaking, both magnetar X-ray bursts and repeating FRBs display similar properties, including a power-law in energy and a log-normal wait time distribution \citep{gogus_1999_apjl,gogus_2000_apjl,gourdji_2019_apjl}. 

Moreover, hyperactive FRBs appear to suddenly switch-on resulting in a large number of FRBs clustered in time, reminiscent of burst storms of magnetars \citep[e.g.,][]{younes_2020}. These properties are broadly consistent with the behaviour expected to emerge from non-linear dynamic systems governed by self-organised criticality (SOC; \citealt{bak_1987}). SOC is widely thought to be applicable to highly magnetized neutron stars due to the separation of timescales of slowly driven crustal stresses due to Hall drift, Ohmic decay, and ambipolar diffusion (e.g., \citealt{1992ApJ...395..250G,pons_vigano}) which occur on relatively long timescales ($\tau > 10^{3} \, {\rm yr}$), and the dissipation in magnetic reconnection or crustal yielding events which may power magnetar X-ray bursts and possibly FRBs ($\tau < 1 \, {\rm sec}$). Minor crustal events triggered by critical crustal stresses may trigger neighbouring regions with near-critical stress \citep{lander_23}, resulting in scale-invariant behaviour and power-law distribution in burst energies predicted by SOC. \\

High-energy magnetar bursts are often loosely categorised by energy radiated in X-/$\gamma$-rays, which is thought to be reasonable proxy for the bolometric energy output. The most common, but least energetic type are the short X-ray bursts ($\sim$$10^{38-41}$\,erg), followed by intermediate flares ($\sim$$10^{41-42}$\,erg), and finally magnetar giant flares (MGFs; $\sim$$10^{45-47}$\,erg). MGFs, the most energetic transient events known from magnetars, are particularly rare occurences in which it is thought that a large fraction of the magnetospheric energy is liberated \citep{thompson_duncan_1995}. In the past half century, three such events have been seen from magnetars in the Milky Way and the Large Magellanic Cloud \citep{mazets_1979,hurley_1999,palmer_2005}, but an extragalactic population is emerging, hidden amongst the population of cataclysmic short/long gamma-ray bursts \citep{burns_2021,beniamini_2025}. \cite{tendulkar_2016_apj} were able to place a constraining upper limit on FRB-like emission associated with the MGF from SGR~1806$-$20, implying if FRBs are co-produced with MGFs beaming must play an important role. \\
As aforementioned, observations thus far suggest a redshift-independent maximum observed (isotropic-equivalent) FRB energy of $\sim$$10^{42}$\,erg for sources at $z \lesssim 2$. While the unknown beaming fraction makes direct comparisons difficult, these most energetic FRBs could plausibly be associated with MGFs given a reasonable radio efficiency factor of $\eta_{\rm r} \sim 10^{-4}$. {\it XMM-Newton} observations were conducted for \ronefourseven by \cite{eppel_2025}, with simultaneous coverage of 459 FRBs detected using the Effelsberg radio telescope. These observations constrain the radio/X-ray fluence ratio for the brightest bursts observed by Effelsberg from \ronefourseven to between $\eta_{\rm r/x} > 4 \times 10^{-7}$ and $\eta_{\rm r/x} > 8 \times 10^{-8}$ depending on the spectral model. These stringent limits are consistent with the possibility that MGFs are co-produced with the brightest FRBs, theoretical predictions of X-ray counterparts \citep{metzger_2019_mnras,margalit_2020_apjl}, and observations of the X-ray counterpart to the FRB-like burst from \sgr \citep{tavani_2021_natas,ridnaia_2021_natas,mereghetti_2020_apjl,li_2021_natas,bochenek_2020_natur,chime_2020_natur_galacticfrb}.\\
High-cadence observations of SGR~1935+2154 have revealed that its radio bursts span $7-8$ orders-of-magnitude in energy \citep{kirsten_2021_natas}. If magnetars are indeed the sources of (most) repeating FRBs, then it is plausible that the observed low-energy ($\sim10^{36-40}$\,erg) and high-energy ($\sim$$10^{40-42}$\,erg) radio bursts trace the intermediate and giant X-ray/$\gamma$-ray flares that are seen from Galactic sources, but which are too weak to detect beyond tens of Mpc extragalactic distance. The broken power-law distribution of burst energies for hyperactive repeaters, with $E_{\rm break} \sim 10^{39-40}$\,erg (Table~\ref{tab:repeater-breaks}), is consistent with this hypothesis because it suggests a transition in burst type and rate --- a transition that may also map the apparent dichotomy between repeaters and apparently one-off bursts, as discussed by \citet{kirsten_2024_natas} and \citet{ouldboukattine_2026_mnras_2}. Likewise, our finding that a large fraction of the source's energy budget can be expended in a single burst (the STROOP), equivalent to many thousands of lower-energy events, is comparable to the energy ratio between MGFs and intermediate flares.

Further evidence for this hypothesis, that FRBs become qualitatively different towards and above high energies of $\sim$$10^{40}$\,erg, can come from finding differences in the spectro-temporal and polarimetric properties of low-/high-energy FRBs. \citet{hewitt_2022_mnras} show that there appears to be a change in burst temporal width and bandwidth between the low-/high-energy bursts seen from FRB~20121102A, mirroring the dichotomy between repeating and apparently non-repeating FRB sources \citep{pleunis_2021_apj}. Also, \citet{hewitt_2023_mnras} find dense `forests' of microshots in some high-energy bursts, which hint at a differing emission mechanism and/or trigger. High-cadence (hundreds to thousands of hours on-source), wide-band ($500-1000$\,MHz), high-time-resolution (microsecond), and polarimetric observations are needed to better quantify the emerging evidence for qualitatively different FRB types at low/high radio energy.

\section{Conclusions} \label{sec:conclusion}
In this paper we present $4{,}233.58$\,hours of monitoring observations of the highly active repeater \ronefourseven as part of the HyperFlash (Onsala, Stockert, \torun, Westerbork, and Dwingeloo) and \'ECLAT (\nancay) campaigns, focusing on the energetics and time-variable DM of the source. To our knowledge, this is the most intensive observing campaign ever targeting a repeating FRB source. We find the following:

\begin{itemize}
    \item We observed $178$~HyperFlash-detected bursts from \ronefourseven, $41$ of which were detected by multiple telescopes: $176$ at L-band ($\sim$$1.4$\,GHz), $2$ at P-band ($\sim$$350$\,MHz), and none at C-band ($\sim$$5$\,GHz). The L-band bursts have a cumulative energy of $4.4 \times 10^{42}$\,erg, assuming isotropic emission and a $1$\,GHz emission bandwidth (as typically assumed in the literature for comparison between studies). This is $\sim$$2\times$ more energy than found in $11{,}553$~bursts detected by FAST \citep{zhang_2025_arxiv}. We thus find that the highest-energy bursts from \ronefourseven have a major and possibly dominant contribution to the depletion of the source's energy budget.
    \item Our most energetic HyperFlash burst, which we term the STROOP, has an isotropic energy of $1.4 \times 10^{42}$\,erg, roughly a third the cumulative energy of our entire $176$ L-band burst sample, and equivalent to roughly $\sim$$11{,}000$ lower-energy bursts from the FAST burst sample. Hence, this underlines that even single events can play a major role in exhausting the \ronefourseven's energy budget. For an assumed radio efficiency of $\eta_{\textrm{r}} \sim 10^{-4}$, the implied total bolometric energy of the STROOP is comparable to the highest-energy magnetar giant flares \citep{kaspi_2017_araa}.
    \item Bursts like the STROOP are exceptionally rare: we monitored \ronefourseven for $17$\,\% of all time ($\sim$$2{,}000$\,hr) during a 480-day ($\sim$$11{,}400$\,hr) period bracketing this brilliant event. While it is of course possible that we were not observing at the times of other similarly energetic events to the STROOP, our observations suggest that even for such an active repeater, STROOP-like events occur at a rate of order one event per $10^3$\,hr.
    \item Using bursts from both the HyperFlash and \eclat observing campaigns, we find that the cumulative burst energy distribution follows a $\gamma = -1.45 \pm 0.02 \pm 0.06$ power-law between $4 \times 10^{29}$\,erg~Hz$^{-1}$ to $2 \times 10^{31}$\,erg~Hz$^{-1}$ and a $\gamma = -0.88 \pm 0.08 \pm 0.13$ power-law between $2 \times 10^{31}$\,erg~Hz$^{-1}$ to $2 \times 10^{33}$\,erg~Hz$^{-1}$. This is qualitatively and roughly quantitatively similar to the high-energy burst distributions seen for other highly active repeaters and emphasises a possible connection between high-energy repeater bursts and the population of apparently one-off FRBs \citep[e.g.,][]{kirsten_2024_natas,ouldboukattine_2026_mnras_2}.
    \item A sub-sample of our HyperFlash bursts have sufficient S/N and fine temporal structure to accurately constrain their DM at the $\sim$$0.1-0.5$\,\dmunit level. Over a close-to 1-year span, we see a $+0.96 \pm 0.06$\,\dmunit linear increase in the DM. During this time, however, the RM remained roughly stable, and the estimate for the parallel magnetic field strength in this time period of $64$\,\micro G is less compared to those of other repeating FRBs \citep[e.g.,][]{ouldboukattine_2026_mnras_2, annathomas_2023_sci}. Although \cite{wang_2026_apjs} showed that the RM decreased later in \ronefourseven's period of activity, we lack sufficiently accurate, contemporaneous DM measurements to further constrain the parallel magnetic field strength. The presence of varying DM and RM values does, however, align with the interpretation that some repeating sources reside in dense and turbulent magnetospheric environments.
\end{itemize}

Overall, our results are consistent with a magnetar origin for \ronefourseven, and we hypothesise that the low-energy ($\sim$$10^{36-40}$\,erg) and high-energy ($\sim$$10^{40-42}$\,erg) radio bursts may trace the intermediate and giant X-ray/$\gamma$-ray flares that are seen from Galactic and nearby extragalactic magnetars. However, \ronefourseven's extreme burst activity and cumulative energy output are beyond what has been observed to date from Galactic magnetars, and could imply a more exotic origin \citep{margalit_2020_apjl}.

\clearpage

\section*{Acknowledgements}
{
We thank Andrew~Jameson for insightful discussions and valuable input on the Heimdall S/N determination, particularly his suggestion regarding the maximum boxcar width.
We thank Alice~Curtin, Ersin~G\"o\u{g}\"u\c{s}, and Daniela~Huppenkothen for insightful discussions.
We thank the directors and staff of the participating telescopes for allowing us to observe with their facilities. 
The AstroFlash research group at McGill University, University of Amsterdam, ASTRON, and JIVE is supported by: a Canada Excellence Research Chair in Transient Astrophysics (CERC-2022-00009); an Advanced Grant from the European Research Council (ERC) under the European Union’s Horizon 2020 research and innovation programme (`EuroFlash’; Grant agreement No. 101098079); an NWO-Vici grant (`AstroFlash’; VI.C.192.045); an NSERC Discovery Grant (RGPIN-2025-06681); an ERC Starting Grant (`EnviroFlash’; Grant agreement No. 101223057); and an NWO-Veni grant (VI.Veni.222.295).
A.~J.~C. acknowledges support from the Oxford Hintze Centre for Astrophysical Surveys which is funded through generous support from the Hintze Family Charitable Foundation.
A.M.C. is a Banting Postdoctoral Fellow.
This work makes use of data from the Westerbork Synthesis Radio Telescope and the Dwingeloo Radio Telescope, both owned by ASTRON. ASTRON, the Netherlands Institute for Radio Astronomy, is an institute of the Dutch Scientific Research Council NWO (Nederlandse Organisatie voor Wetenschappelijk Onderzoek). We thank the Westerbork operators Richard Blaauw, Jurjen Sluman, and Henk Mulder for scheduling and supporting observations. 
F.~K. acknowledges support from Onsala Space Observatory for the provisioning of its facilities/observational support. The Onsala Space Observatory national research infrastructure is funded through Swedish Research Council grant No 2017-00648.
We express our gratitude to the operators and observers of the Astropeiler Stockert telescope: Thomas Buchsteiner, Elke Fischer, and Hans-Peter L\"oge.
We thank the operators of the Dwingeloo radio telescope: Tjipke de Beer, Simon Bijlsma, Gerard Boons, Paul Boven, Hans van der Meer, Harm Munk, Roel Ovinge, Michel Sanders, Marc Wolf, and all CAMRAS-volunteers who keep the telescope operational.
This work is based in part on observations carried out using the 32-m radio telescope operated by the Institute of Astronomy of the Nicolaus Copernicus University in \torun (Poland) and supported by a Polish Ministry of Science and Higher Education SpUB grant. 
The \nancay Radio Observatory is operated by the Paris Observatory, associated with the French {\it Centre National de la Recherche Scientifique} (CNRS). We acknowledge financial support from the {\it Programme National de Cosmologie et Galaxies} (PNCG) and {\it Programme National Hautes Energies} (PNHE) of INSU, CNRS, France.
}


\section*{Data Availability}
The data that support the plots within this paper, burst snippets of the STROOP, and other findings of this study are available under \url{https://doi.org/10.5281/zenodo.20185181} or from the corresponding author upon reasonable request.
The scripts and Jupyter notebooks used to analyse the data, generate the plots and tables with the burst properties are available at \url{https://github.com/astroflash-frb/frb20240114a-ouldboukattine-2026}. \\

The FRB software pipeline, \texttt{FRB-baseband}, written to process and search the baseband data for the stations Westerbork, Onsala and \torun can be found at \url{https://github.com/pharaofranz/frb-baseband}. The specific SFXC version to create the coherently dedispersed filterbank files, \texttt{SFXC-phased-array}, is hosted at \url{https://github.com/aardk/sfxc/tree/phased-array}. 
The Dwingeloo FRB software pipeline, \texttt{Dwingeloo-frbscripts}, can be found at \url{https://gitlab.camras.nl/dijkema/frbscripts/}. \texttt{vrt-iq-tools} can be retrieved from \url{https://github.com/tftelkamp/vrt-iq-tools}
The NRT analyzer python scripts, \texttt{ECLAT-burst-analyzer}, can be found at \url{https://github.com/astroflash-frb/burst_analyzer}. The software to mitigate RFI, \texttt{jess}, can be found at \url{https://github.com/josephwkania/jess}.
{\tt jive5ab} can be found on \url{https://github.com/jive-vlbi/jive5ab}, {\tt Heimdall} is hosted at \url{https://sourceforge.net/projects/heimdall-astro/} and \texttt{FETCH} can be found at \url{https://github.com/devanshkv/fetch}. 
The pulsar package {\tt DSPSR} is hosted at \url{https://sourceforge.net/projects/dspsr/} and {\tt SIGPROC} can be retrieved from \url{https://github.com/SixByNine/sigproc}.

This work made use of the following software packages: \texttt{astropy} \citep{astropy:2013,astropy:2018,astropy:2022}, \texttt{Jupyter} \citep{2007CSE.....9c..21P,kluyver2016jupyter}, \texttt{matplotlib} \citep{Hunter:2007}, \texttt{numpy} \citep{numpy}, \texttt{pandas} \citep{mckinney-proc-scipy-2010}, \texttt{python} \citep{python}, \texttt{scipy} \citep{2020SciPy-NMeth}, and \texttt{tqdm} \citep{tqdm_14231923}.
This research has made use of the Astrophysics Data System, funded by NASA under Cooperative Agreement 80NSSC21M00561.
Software citation information aggregated using \texttt{\href{https://www.tomwagg.com/software-citation-station/}{The Software Citation Station}} \citep{software-citation-station-paper,software-citation-station-zenodo}.


\bibliographystyle{mnras}
\bibliography{references} 



\appendix

\bsp	
\clearpage
\label{lastpage}

\section{Calculations related to the break in the energy distribution}
\label{app:interpretation_break_maths}
In the following, we describe how different distributions in beaming and the true energy of bursts affects the observed (isotropic equivalent) cumulative rate distribution. This section supplements Section \ref{sect:interpretation_break} in the main text, where we present the primary conclusions derived from the analysis below. 

\subsection{Case 1: Fixed \texorpdfstring{$E_{\rm true}$}{E true}, Power-law \texorpdfstring{$p(f_{\rm b})$}{p(f b)}}
Let $p(f_{\rm b})$, probability of an FRB to be emitted with a beam factor $f_{\rm b}$, take a power-law form such that $p(f_{\rm b}) \propto f_{\rm b}^{-\alpha}$, bounded by $f_{\rm b, min} \leq f_{\rm b} \leq 1$. Then the probability of detecting a burst as a function of $f_{\rm b}$ is weighted:
\begin{equation}
    \frac{d\dot{N}_{\rm obs}}{d f_{\rm b}} \propto p_{\rm obs}(f_b) \propto f_b p(f_{\rm b}) \propto f_b^{1-\alpha}
\end{equation}
As $E_{\rm obs} = E_{\rm true}/f_{\rm b}$, then $|df_{\rm b}/d E_{\rm obs}| = E_{\rm true}/E_{\rm obs}^2$ such that we can derive a rate of FRBs observed as a function of observed (isotropic) energy:
\begin{equation}
\begin{split}
 \frac{d \dot{N}}{d E_{\rm obs}} &\propto p_{\rm obs}(f_{\rm b}) \frac{d f_{\rm b}}{d E_{\rm obs}} \\
 &\propto \bigg(\frac{E_{\rm true}}{E_{\rm obs}}\bigg)^{1 - \alpha} \frac{E_{\rm true}}{E_{\rm obs}^2} \propto E_{\rm obs}^{-(3- \alpha)} 
 \end{split}
\end{equation}
To reproduce observed distributions where $\frac{d \dot{N}}{d E_{\rm obs}} \propto E_{\rm obs}^{-(0.5-2.)}$, we require values of $1 < \alpha < 2.5$, corresponding to narrowly collimated bursts being less common, with a cut-off present at $E_{\rm obs, max} = E_{\rm true}/f_{\rm b,min}$. 

\subsection{Case 2: Power-law \texorpdfstring{$E_{\rm true}$}{E true}, Power-law \texorpdfstring{$p(f_{\rm b})$}{p(f b)}}
Now let the underlying distribution of true FRB energies be a power-law function of the form: $\Phi(E_{\rm true}) \propto E_{\rm true}^{-\gamma}$, and $p(f_{\rm b}) \propto f_{\rm b}^{-\alpha}$ as before. Now we have, via integration over all beaming fractions:
\begin{equation}
\begin{split}
\frac{d \dot{N}}{d E_{\rm obs}} &\propto \int_{f_{\rm b,min}}^{1} p_{\rm obs}(f_{\rm b}) \Phi(E_{\rm true}) d f_{\rm b} \\ &\propto \int_{f_{\rm b,min}}^{1} f_{\rm b}^{1 - \alpha} (f_{\rm b} E_{\rm obs})^{-\gamma} f_{\rm b} d f_{\rm b} 
&\propto E_{\rm obs}^{-\gamma}
\end{split}
\end{equation}
In this case, the $\frac{d \dot{N}}{d E_{\rm obs}}$ slope has no dependence on $\alpha$, and instead the quantity $\int_{f_{\rm b, min}}^1 f_{\rm b}^{2-\alpha-\gamma} d f_{\rm b}$ sets the normalization of the $\frac{d \dot{N}}{d E_{\rm obs}}$. This means that any break in the power-law distribution of the FRB energy rate distribution cannot be explained by a power-law in $p(f_{\rm b})$ and in $E_{\rm true}$.

\subsection{Case 3: Power-law \texorpdfstring{$E_{\rm true}$}{E true}, Broken Power-law \texorpdfstring{$p(f_{\rm b})$}{p(f b)}}
To understand where the observed break comes from, we allow $p(f_{\rm b})$ distribution to take the form of a broken power-law such that:
\begin{equation}
p(f_{\rm b}) \propto
\begin{cases}
f_{\rm b}^{-\alpha_1}, & f_{\rm b,min} < f_{\rm b} < f_{\rm b,crit} \\
f_{\rm b}^{-\alpha_2}, & f_{\rm b,crit} < f_{\rm b} < 1
\end{cases}
\label{app_eq:case3_bpl}
\end{equation}
In this case, the observed rate energy distribution is:
\begin{equation}
\begin{split}
    \frac{d \dot{N}}{d E_{\rm obs}} &\propto E_{\rm obs}^{-\gamma} \bigg( \int_{f_{\rm b,min}}^{\rm f_{\rm b, crit}} f_{\rm b}^{2 - \alpha_1 - \gamma} d f_{\rm b} +  \int_{f_{\rm b, crit}}^{1} f_{\rm b}^{2 - \alpha_2 - \gamma} d f_{\rm b} \bigg)  &\propto E_{\rm obs}^{-\gamma} \\    
\end{split}
\end{equation}
Once again, we find that a broken power-law in $p(f_{\rm b})$ cannot explain observed break in $\frac{d \dot{N}}{d E_{\rm obs}}$ of FRBs.

\subsection{Case 4: Broken Power-law \texorpdfstring{$E_{\rm true}$}{E true}, Power-law \texorpdfstring{$p(f_{\rm b})$}{p(f b)}}
Finally, we consider the case where the underlying distribution of the true FRB energy takes the form of a broken power-law:
\begin{equation}
\Phi(E_{\rm true}) \propto
\begin{cases}
E_{\rm true}^{-\gamma_1}, & E_{\rm true} < E_{\rm true, crit} \\
E_{\rm true}^{-\gamma_2}, & E_{\rm true} > E_{\rm true, crit}
\end{cases}
\label{app_eq:case4_bpl_true}
\end{equation}
In this case:
\begin{equation}
\frac{d \dot{N}}{d E_{\rm obs}} \propto \int_{f_{\rm b,min}}^{1} \Phi(f_{\rm b} E_{\rm obs}) f_{\rm b}^{2-\alpha} d f_{\rm b} 
\end{equation}
This can be evaluated in three regimes, reducing to two limiting regimes when $f_{\rm b, min} \rightarrow 0$.
:
\begin{equation}
\begin{aligned}
\frac{d \dot{N}}{d E_{\rm obs}} \propto
\begin{cases}
E_{\rm obs}^{-\gamma_1}, & E_{\rm obs} < E_{\rm true, crit} \\
C_1 E_{\rm obs}^{-(3-\alpha)} + C_2 E_{\rm obs}^{-\gamma_1} + C_3 E_{\rm obs}^{\gamma_2}, 
& E_{\rm true, crit} < E_{\rm obs} < \frac{E_{\rm true, crit}}{f_{\rm b, min}} \\
E_{\rm obs}^{-\gamma_2}, & E_{\rm obs} > \frac{E_{\rm true, crit}}{f_{\rm b,min}}
\end{cases}
\end{aligned}
\label{app_eq:case4_bpl_observed}
\end{equation}
Here, the total rate-energy distribution consists of the low-energy power-law ($E_{\rm obs}^{-\gamma_1}$), a transition region where integration constants set relative power-law weights, and a high-energy power-law ($E_{\rm obs}^{-\gamma_2}$). The transition region spans roughly $1/f_{\rm b, min}$ in energy (e.g., 1 decade in energy for $f_{\rm b, min} = 0.1$), the final regime is never reached for $f_{\rm b, min} \rightarrow 0$. In the case that $f_{\rm b, min} \rightarrow 1$, corresponding to all bursts being isotropic, only regimes 1 and 3 are relevant, matching the underlying distribution of $\Phi(E_{\rm true})$ as expected. A summary of our conclusions is presented in the main text in Section \ref{sect:interpretation_break}.

\section{Supplementary figures and tables}

\begin{figure*}
    \centering
    \includegraphics[width=\linewidth]{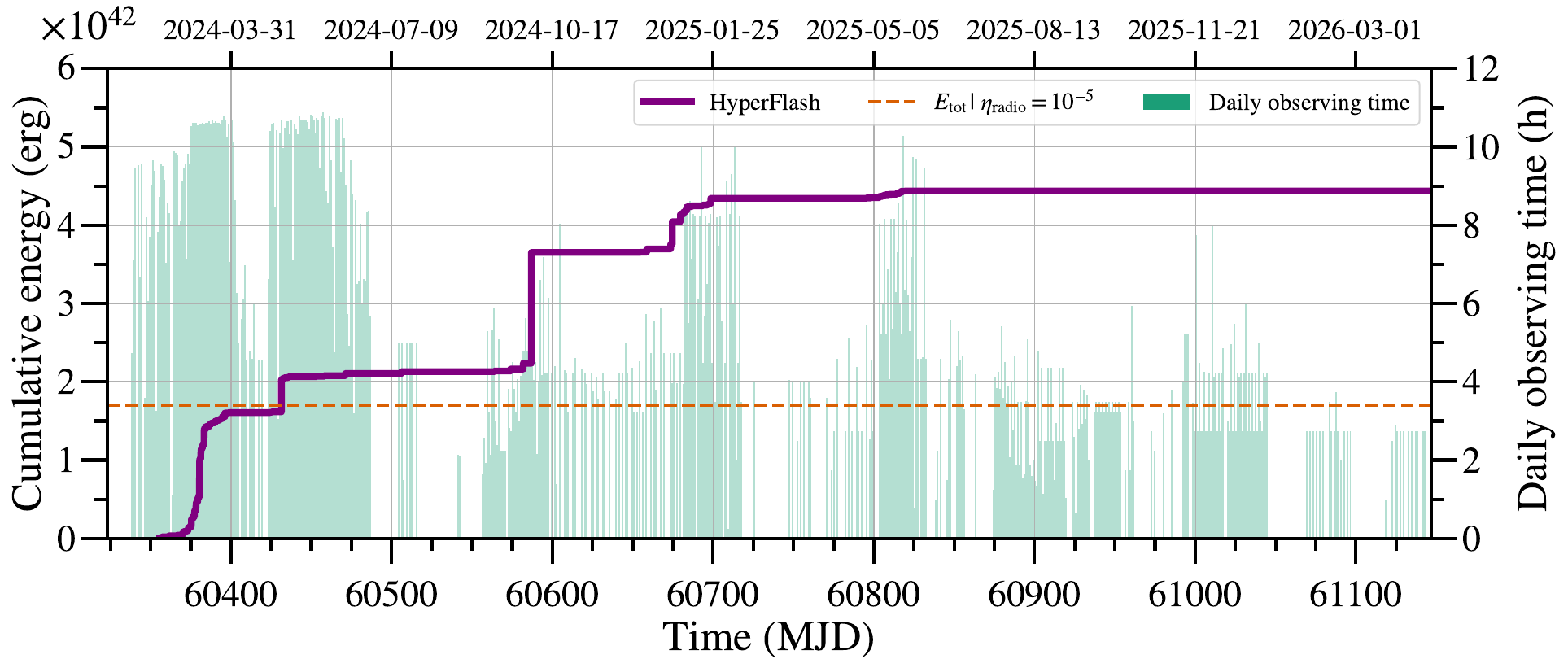}
    \caption{Cumulative sum of burst energies for the HyperFlash sample (purple), with energies scaled to a $1$\,GHz emission bandwidth. The light green histogram shows the daily exposure towards \ronefourseven, accounting for overlap between the HyperFlash stations. Due to the low declination of \ronefourseven($\sim$\,$+4\degree$) and the northern latitudes of the HyperFlash telescopes, the maximum exposure is $\sim$$11.2$\,hours per day. The orange dotted line indicates the estimated stored magnetic energy of a magnetar, assuming a dipolar field and a radio efficiency of $\eta_{\mathrm{r}} = 10^{-5}$, at $1.7\times10^{42}$\,erg.}
    \label{fig:cumulative_sum_energy_hf}
\end{figure*}

\begin{figure*}
    \centering
    \includegraphics[width=\linewidth]{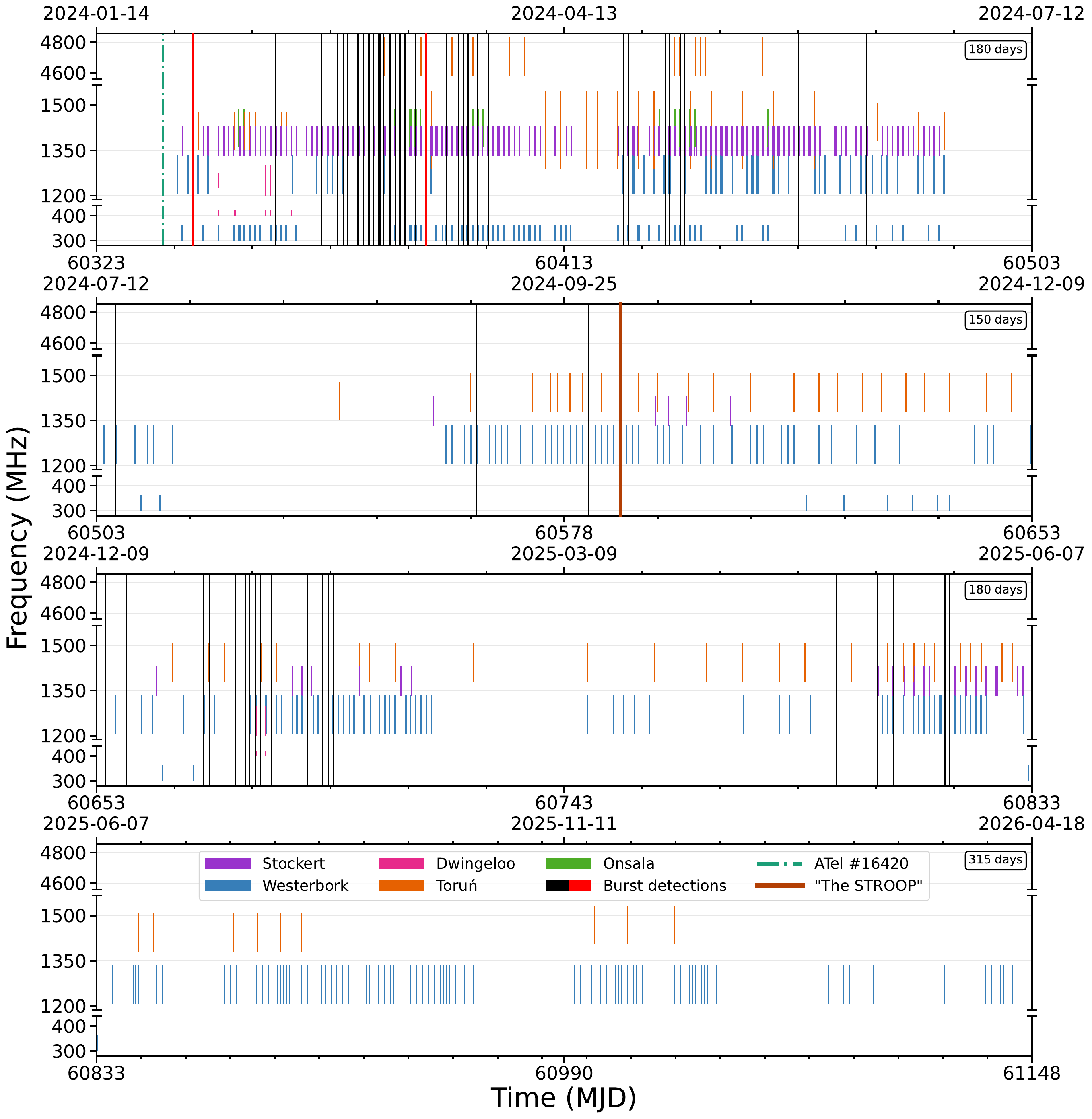}
    \caption{Overview of the HyperFlash observational campaign targeting \ronefourseven. Each coloured block represents an observation with a specific telescope, where the height indicates the covered frequency range and the width the observing duration. Observations were carried out at C-band ($\sim$$4.8$\,GHz), L-band ($\sim$$1.4$\,GHz), and P-band ($\sim$$324$\,MHz), with durations ranging from $1$ to $11.2$\,hours. Due to the source elevation and the high latitudes of the HyperFlash telescopes located in northern Europe, \ronefourseven was observable for at most $11.2$\,hours per day. The observing campaign spanned $806$\,days in total and is divided into panels ranging between $150$ and $315$\,days, as indicated in the top right of each panel. Burst detections are marked by vertical lines: black for L-band and red for P-band detections; no detections were made at C-band. For visual clarity, only a subset of detections in February and March 2024 is shown. The discovery ATel \citep{shin_2024_atel} is indicated by a dash-dotted green line. We also highlight our brightest detection, the STROOP, shown in Figure \ref{fig:the_stroop}, with a vertical caramel-coloured line.}
    \label{fig:obs_campaign_r147}
\end{figure*}

\begin{figure*}
    \centering
    \includegraphics[width=0.98\linewidth]{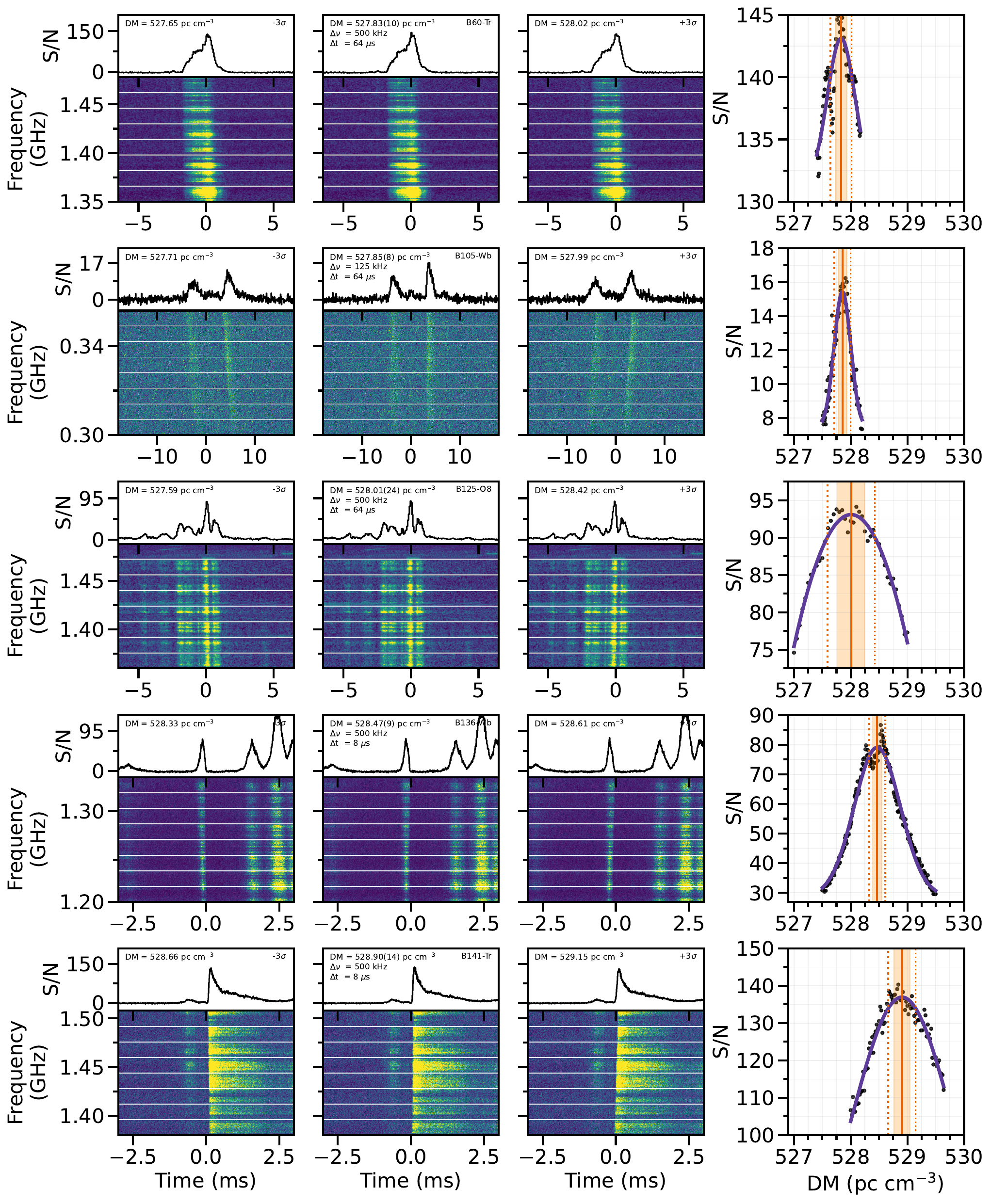}
    \caption{Dynamic spectra (bottom panels), time series (top panels), and DM vs. peak S/N curves for five bursts that exhibit fine ($\lesssim100$\,\microsec) temporal structure, enabling accurate DM measurements. We show five of the thirteen bursts with DM measurements presented in Figure~\ref{fig:dm_evolution}. For each burst, the second column shows the data at the S/N-optimized DM, while the first and third columns show the burst at $\pm 3\sigma$. All bursts have been coherently and incoherently dedispersed to the indicated DM. We zoom in on the specific burst component used for the DM optimization. The fourth column shows Gaussian fits to the S/N versus DM curves (purple). The solid orange line indicates the best-fit DM, the shaded orange region the $1\sigma$ uncertainty, and the dotted orange lines the $3\sigma$ range corresponding to the DMs shown in the first and third columns.}
    \label{fig:r147_dynamic_matrix}
\end{figure*}

\begin{figure}
    \centering
    \includegraphics[width=\columnwidth]{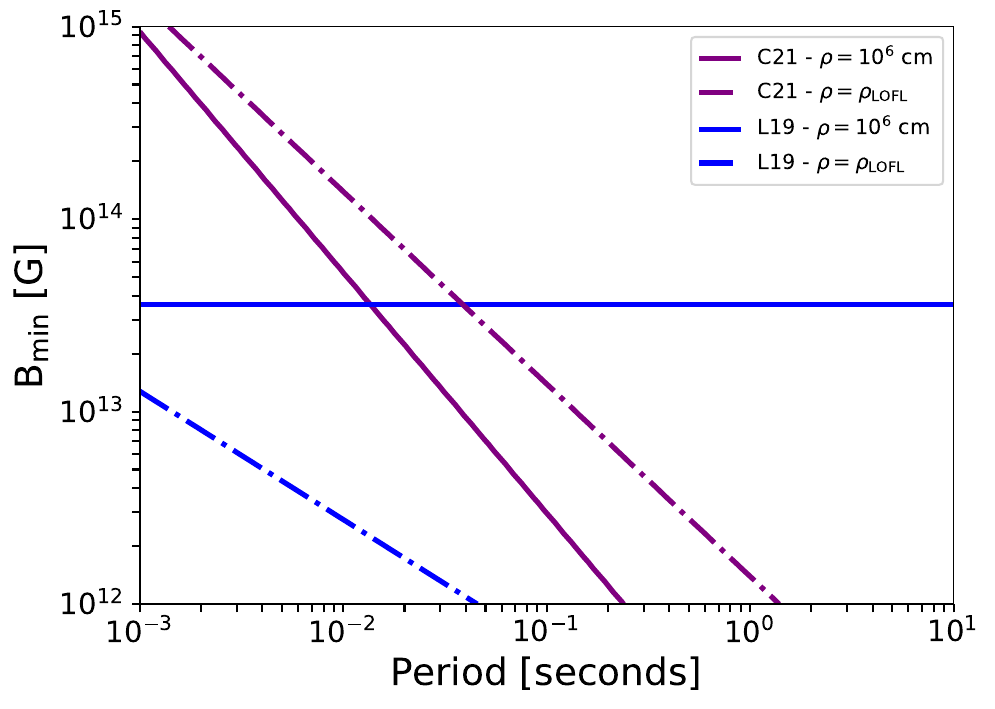}
    \caption{Minimum local magnetic field constraints as a function of neutron star rotation period for the STROOP due to Schwinger pair production \protect\citep{lu_maximum_2019} and momenta alignment \protect\citep{cooper_2021_mnras}, both assuming coherent curvature radiation. Solid lines refer to fixed field line curvature radius of $\rho_{\rm c} = 10^{7} \, {\rm cm}$, dot-dashed lines refer to emission along the last open field line (LOFL) at the polar cap co-latitude.}
    \label{fig:ccr_magnetospheric_limits}
\end{figure}

Appendix Figure~\ref{fig:cumulative_sum_energy_hf} shows the cumulative sum of burst energies for the HyperFlash sample, along with the daily exposure of the five participating telescopes, accounting for overlap between stations, 
while Appendix Figure~\ref{fig:obs_campaign_r147} provides an observational overview of the HyperFlash campaign on \ronefourseven. 
Appendix Table~\ref{tab:evolving_dm} lists DM measurements for a subsample of bursts for which accurate determinations were possible. 
Appendix Table~\ref{tab:obs_coverage} summarizes the observational setups of the different observing modes, including detections, completeness thresholds, and total observing time per telescope and mode. 
In Appendix Table~\ref{tab:hyperflash_burst_prop_appendix}, we present the burst properties detected during the HyperFlash campaign; the full dataset is also available in \texttt{.csv} format as Supplementary material, on the provided GitHub repository, and on Zenodo. 
Appendix Figure~\ref{fig:r147_dynamic_matrix} presents the dynamic spectra, time series, and Gaussian-fitted DM versus peak S/N curves for five bursts, supporting Figure~\ref{fig:dm_evolution}, which shows a linear increase of DM with time. 
Finally, Appendix Figure~\ref{fig:ccr_magnetospheric_limits} presents the minimum local magnetic field for the STROOP assuming coherent curvature radiation. These constraints arise due to Schwinger pair production (from Equation \ref{eq:lu_constraints}; \citealt{lu_maximum_2019}) and Equation~15 in \citet{cooper_2021_mnras}. These minimum magnetic field constraints are shown as a function of neutron star period for both fixed field line curvature radius $\rho = 10^{6} \, {\rm cm}$ and  $\rho_{\rm LOFL}$, the curvature radius of the last open field line in the polar cap region.

\begin{table}
\caption{DM measurements were obtained for a subsample of bursts for which accurate estimates were possible. The corresponding linear fit is shown in Figure~\ref{fig:dm_evolution}, while the dynamic spectra of five selected bursts, illustrating the DM increase, are displayed in Appendix~Figure~\ref{fig:r147_dynamic_matrix}.}
\label{tab:evolving_dm}
\resizebox{\columnwidth}{!}{%
\begin{tabular}{@{}c c S[table-format=5.5,group-digits=none] S[table-format=3.3,group-digits=none] S[table-format=1.3,group-digits=none]@{}}
\hline
\hline
{Burst ID} & {Station} & {ToA}         & {DM}                 & {DM-error} \\
                    & & {[MJD]}            & {[\dmunit]}          & {[\dmunit]}                  \\ \hline
B57                 & O8 & 60380.27609 & 528.085            & 0.508                               \\
B60                 & Tr & 60380.43194 & 527.830            & 0.102                               \\
B63                 & O8 & 60380.45214 & 528.146            & 0.160                               \\
B102                & O8 & 60383.26844 & 528.205            & 0.260                               \\
B105                & Wb & 60386.37677 & 527.854            & 0.075                               \\
B125                & O8 & 60431.43777 & 528.007            & 0.240                               \\
B135                & Wb & 60573.92099 & 528.313            & 0.103                               \\
B136                & Wb & 60581.87250 & 528.464            & 0.086                               \\
B137$^\mathrm{a}$   & Wb & 60586.95299 & 528.468            & 0.081                               \\
B140                & Wb & 60673.60990 & 528.761            & 0.406                               \\
B141                & Tr & 60674.67263 & 528.904            & 0.139                              \\
B158                & Tr & 60686.62063 & 528.746            & 0.471                               \\
B163                & Tr & 60698.53842 & 528.673            & 0.560                               \\ \bottomrule
\multicolumn{5}{l}{$\mathrm{^{a}}$"The STROOP" shown in Figure~\ref{fig:the_stroop}} \\
\end{tabular}%
}
\end{table}

\begin{table*} 
\caption{\label{tab:obs_coverage}\textbf{Observational set-up}}
\resizebox{\textwidth}{!}{%
\begin{tabular}{c c c c c c S[table-format=2.1] S[table-format=3.1] S[table-format=3.2,input-decimal-markers={.,}]}
\hline
\hline
{Station$\mathrm{^{a}}$}  & {Band} & {Frequency} & {Bandwidth$\mathrm{^{b}}$} & {Bandwidth per} & {SEFD} & {Detection$\mathrm{^{d}}$} & {Completeness$\mathrm{^{e}}$} & {Time observed$\mathrm{^{f}}$} \\
 &  & {[MHz]} & {[MHz]} & {subband [MHz]} & {[Jy]} & {threshold [Jy~ms]} & {threshold [Jy~ms]} & {[hr]} \\
\hline
Wb  & P$_{\rm Wb}$      & 300--356          &50     & 8   & 2100$\mathrm{^{c}}$     & 46.5                   & 172.5                  & 676.72 \\
Dw  & P$_{\rm DW}$      & 400--420          &20     & 20  & $\sim 2100$             & \multicolumn{1}{c}{--} & \multicolumn{1}{c}{--} & 27.55 \\
Dw  & L$_{\rm Dw}$      & 1200--1400        &180    & 100 & $\sim 850$              & \multicolumn{1}{c}{--} & \multicolumn{1}{c}{--} & 20.52 \\
Wb  & L$_{\rm Wb}$      & 1207--1335        &100    & 16  & 420$\mathrm{^{c}}$      & 6.6                    & 24.4                   & 1539.84 \\
NRT & L$_{\rm NRT}$     & 1228--1740        &500    & 512 & 25                      & 0.18                   & 0.65                   & 55.99$\mathrm{^{g}}$ \\
Tr  & L$_{\rm Tr-1}$    & 1290--1546        &200    & 32  & 350$\mathrm{^{c}}$      & 3.9                    & 14.4                   & 69.22 \\
St  & L$_{\rm St}$      & 1332.5--1430.5    &90     & 98  & 385                     & 6.4                    & 23.6                   & 1228.14 \\
Tr  & L$_{\rm Tr-2}$    & 1350--1478        &100    & 16  & 350$\mathrm{^{c}}$      & 5.5                    & 20.3                   & 48.17 \\
O8  & L$_{\rm O8}$      & 1360--1488        &100    & 16  & 310$\mathrm{^{c}}$      & 4.8                    & 18.0                   & 167.47 \\
Tr  & L$_{\rm Tr-3}$    & 1380--1508        &200    & 32  & 350$\mathrm{^{c}}$      & 5.5                    & 20.3                   & 303.17 \\
Tr  & L$_{\rm Tr-4}$    & 1405--1533        &100    & 16  & 350$\mathrm{^{c}}$      & 5.5                    & 20.3                   & 32.61 \\
Tr  & C$_{\rm Tr-1}$    & 4580--4836        &200    & 32  & 220$\mathrm{^{c}}$      & 2.4                    & 9.0                    & 64.18 \\
\hline
\multicolumn{8}{l}{Total non-overlapping observing time at L-band for HyperFlash/NRT, used in the computation of left panel of Figure~\ref{fig:burst_distr_r147} [hr]} & \textrm{1253.77/39.70} \\
\multicolumn{8}{l}{Total non-overlapping observing time at L-band for HyperFlash/NRT, used in the computation of right panel of Figure~\ref{fig:burst_distr_r147} [hr]} & \textrm{1826.00/55.99} \\
\hline
\multicolumn{8}{l}{Total non-overlapping time at L-band for all stations$\mathrm{^{f}}$} & \textrm{2688.24} \\
\hline
\multicolumn{8}{l}{Total telescope time/total non-overlapping time on source [hr]$\mathrm{^{f,g}}$} & \textrm{4233.58/2861.53} \\
\hline

\multicolumn{8}{l}{$\mathrm{^{a}}$ Wb: Westerbork RT-1, St: Stockert, Tr: \torun, O8: Onsala, Dw: Dwingeloo and NRT: \nancay Radio Telescope} \\
\multicolumn{8}{l}{$\mathrm{^{b}}$ Effective bandwidth accounting for RFI and band edges.} \\
\multicolumn{8}{l}{$\mathrm{^{c}}$ From the \href{https://www.evlbi.org/sites/default/files/shared/EVNstatus.txt}{EVN status page}.} \\
\multicolumn{8}{l}{$\mathrm{^{d}}$ Assuming a $7\sigma$ detection threshold and a FRB pulse width of $1~\mathrm{ms}$.} \\
\multicolumn{8}{l}{$\mathrm{^{e}}$ Assuming a $15\sigma$ detection threshold and a width of $3~\mathrm{ms}$.} \\
\multicolumn{8}{l}{$\mathrm{^{f}}$ Total on-source observing hours over the full observational campaign spanning MJD $60337$--$61143$.} \\
\multicolumn{8}{l}{$\mathrm{^{g}}$ Total on-source observing hours over the full NRT observational campaign spanning MJD $60344$--$60723$.} \\
\end{tabular}
}
\end{table*}

\begin{table*}
\caption{\textbf{Burst properties for bursts detected. The full table is available in \texttt{.csv} format in the Supplementary Material.}}
\label{tab:hyperflash_burst_prop_appendix}
\resizebox{\textwidth}{!}{%
\begin{tabular}{c c S[table-format=5.6,group-digits=none] S[table-format=1.2] S[table-format=3.2(4),separate-uncertainty=true] S[table-format=2.2] S[table-format=2.2(4),separate-uncertainty=true] S[table-format=2.2(3),separate-uncertainty=true] c c}
\hline
\hline
{Burst ID$^{\dagger}$} & Station & {TOA$^\mathrm{a}$} & {Peak S/N} & {Fluence$^\mathrm{b}$} & {Width} & {Spectral density$^\mathrm{c}$} & {Spectral luminosity$^\mathrm{d}$} & BW$^\mathrm{e}$ & Central Frequency \\
 &  & {[MJD]} &  & {[Jy ms]} & [ms] & {[$\mathrm{10^{30}\,erg\,Hz^{-1}}$]} & {[$\mathrm{10^{32}\,erg\,s^{-1}\,Hz^{-1}}$]} & [MHz] & [MHz] \\ \midrule

B01 & Wb & 60341.533745 & 8.08 & 203.23 \pm 40.65 & 45.06 & 72.22 \pm 14.44 & 16.03 \pm 3.21 & 40 & 328 \\
B02 & Wb & 60355.639527 & 5.26 & 16.66 \pm 3.33 & 14.34 & 5.92 \pm 1.18 & 4.13 \pm 0.83 & 128 & 1271 \\
B03 & St & 60357.339134 & 8.60 & 20.15 \pm 4.03 & 3.71 & 7.16 \pm 1.43 & 19.28 \pm 3.86 & 98 & 1381 \\
B04 & St & 60357.433876 & 6.00 & 25.10 \pm 5.02 & 9.61 & 8.92 \pm 1.78 & 9.28 \pm 1.86 & 53 & 1381 \\
B05 & St & 60361.542887 & 5.21 & 35.49 \pm 7.10 & 7.86 & 12.61 \pm 2.52 & 16.04 \pm 3.21 & 98 & 1381 \\
B06 & St & 60366.341592 & 5.26 & 16.72 \pm 3.34 & 3.50 & 5.94 \pm 1.19 & 17.00 \pm 3.40 & 98 & 1381 \\

{$\vdots$} & $\vdots$ & {$\vdots$} & {$\vdots$} & {$\vdots$} & {$\vdots$} & {$\vdots$} & {$\vdots$} & {$\vdots$} & {$\vdots$} \\

B175 & Wb & 60814.157155 & 9.01 & 27.84 \pm 5.57 & 12.29 & 9.90 \pm 1.98 & 8.05 \pm 1.61 & 128 & 1271 \\
B176 & Tr & 60816.177709 & 4.85 & 7.27 \pm 1.45 & 7.17 & 2.58 \pm 0.52 & 3.61 \pm 0.72 & 128 & 1444 \\
B177 & Wb & 60816.333009 & 4.82 & 23.81 \pm 4.76 & 10.75 & 8.46 \pm 1.69 & 7.87 \pm 1.57 & 128 & 1271 \\
B178 & Wb & 60817.072561 & 8.03 & 21.07 \pm 4.21 & 12.29 & 7.49 \pm 1.50 & 6.09 \pm 1.22 & 128 & 1271 \\
B179 & Wb & 60817.072561 & 7.89 & 28.74 \pm 5.75 & 16.38 & 10.21 \pm 2.04 & 6.24 \pm 1.25 & 128 & 1271 \\
B180 & Tr & 60819.332590 & 7.58 & 7.22 \pm 1.44 & 7.17 & 2.57 \pm 0.51 & 3.58 \pm 0.72 & 95 & 1444 \\
\bottomrule
\multicolumn{10}{l}{$\mathrm{^{\dagger}}$We detected $219$ bursts, $41$ of which were observed by multiple telescopes.} \\
\multicolumn{10}{l}{\quad Bursts detected by the same telescope have a different station identifier (e.g. B$30$-Wb and B$30$-St)}. \\ 
\multicolumn{10}{l}{\quad The SPC algorithm could not be applied for B$125$-O8 and B$157$-Tr due to much RFI presence.} \\
\multicolumn{10}{l}{\quad B$69$ and B$111$ are missing because they were later identified as duplicates of B$68$ and B$110$, respectively.} \\
\multicolumn{10}{l}{\quad No burst properties were measured for the Dwingeloo bursts, B$48$-Wb, B$81$-O8, B$86$-Tr and B$88$-Tr because the bursts were deemed to weak.} \\
\multicolumn{10}{l}{$\mathrm{^{a}}$Time of arrival referenced to the solar system barycentre at infinite frequency in TDB.} \\
\multicolumn{10}{l}{\quad We adopt a DM of $527.7$\,\dmunit for all bursts and assume a dispersion constant of $\mathcal{D} =1/(2.41 \times 10^{-4})$\,MHz$^{2}$\,pc$^{-1}$\,cm$^{3}$\,s.} \\
\multicolumn{10}{l}{\quad The reference frequency used for scaling to infinite frequency is burst dependent. Bursts processed with \sfxc are referenced to the centre of the top subband,}\\
\multicolumn{10}{l}{\quad whereas bursts created solely with \digifil (i.e. Stockert and the 2-bit reprocessed Westerbork bursts) are referenced to the centre of the top frequency channel.}\\
\multicolumn{10}{l}{\quad The coordinates are as follows. Onsala:  $X = 3370965.8787$\,m, $Y = 711466.1978$\,m $Z = 5349664.2006$\,m. \torun:  $X = 3638558.5100$\,m, $Y = 1221969.7200$\,m }\\
\multicolumn{10}{l}{\quad $Z = 5077036.7600$\,m. Stockert: $X = 4031510.647$\,m, $Y = 475159.114$\,m and $Z = 4903597.840$\,m. Westerbork: $X = 3828750.6969$\,m, $Y = 442589.2176$\,m and}\\
\multicolumn{10}{l}{\quad  $Z = 5064921.5700$\,m; The locations for Westerbork, \torun and Onsala can also be retrieved on the \href{https://github.com/jive-vlbi/sched/blob/python/catalogs/locations.dat}{EVN station locations} page.}\\
\multicolumn{10}{l}{$\mathrm{^{b}}$We assume a $20\%$ error for all bursts dominated by the uncertainty on the SEFD.} \\
\multicolumn{10}{l}{\quad The SEFD values for Westerbork, Onsala and \torun are documented on \href{https://www.evlbi.org/sites/default/files/shared/EVNstatus.txt}{EVN status page}} \\
\multicolumn{10}{l}{$\mathrm{^{c}}$Computed using Equation \ref{eq:energy_r147}, $D_L=616$~Mpc and $z=0.130287$.} \\
\multicolumn{10}{l}{$\mathrm{^{d}}$Spectral density divided by the width.} \\
\multicolumn{10}{l}{$\mathrm{^{e}}$Measured bandwidth used to compute fluence.} \\
\end{tabular}%
}
\end{table*}

\section{\texttt{Heimdall} search strategy} \label{sec:heimdall_option}

In our burst search strategy, we make use of \texttt{Heimdall} as the burst search engine in the \texttt{FRB-baseband} search pipeline. This pipeline is deployed at three of the five participating telescopes, namely Westerbork, Onsala, and \torun.

In 2023, we commissioned and deployed a new recording and processing machine at Westerbork. While reprocessing bursts from \ronefourseven detected with the Onsala telescope on this new machine, we found that we could not re-detect some of the bursts previously detected using the same pipeline running on the server at Onsala. We traced this discrepancy to a different version of \texttt{Heimdall}: the version at Onsala was built from a version prior to 23 March 2021 (commit \texttt{[798707]}), whereas the version at Westerbork was built from a version after this date. This update (commit \texttt{[798707]}) modified the default behaviour for computing the S/N of a burst candidate. Instead of estimating the S/N from the root mean square (RMS) of the filtered time series and normalizing by the same RMS, the updated method first rescales the filtered time series by the width of the boxcar used in the convolution. As a result, the measured S/N values for burst candidates are more accurate and no longer underestimated, as reported by see \citet{gupta_2021_mnras}, see their Appendix~A for a detailed explanation.

In our testing, we found that some bursts previously detected using the original method of S/N computation were missed when using the updated version. This suggested that bursts might have been missed in the Westerbork dataset that was searched on the new server where the latest version of Heimdall was installed. To investigate this, we retrieved all Westerbork-scans that had targeted \ronefourseven up to 31 August 2025 from our long-term archive. These archived filterbank files had been downsampled to a time resolution of $256$\,\microsec (originally searched at $64$\,\microsec), have a frequency resolution of $31.25$\,kHz (originally $15.625$\,kHz), and are stored as 2-bit data (8-bit originally) to keep the data volume manageable. In total, we recovered $7554$ filterbank files from $324$ observations spanning approximately $17$ months, covering both L- and P-band observations. We reprocessed these filterbanks using  \texttt{Heimdall}'s new \texttt{boxcar-renorm} option, which instructs \texttt{Heimdall} to use the original S/N computation method. Since the filterbank files had been downsampled, the maximum boxcar width increased from an original $65$\,ms to $262$\,ms. After reprocessing, we recovered all $42$ previously identified Westerbork bursts (i.e. those that were found by \texttt{Heimdall} using the new S/N-computation algorithm) and found $20$ additional bursts at L-band. At P-band, no new bursts were found. The \texttt{Heimdall} reported S/N values of these additional bursts range between $7.2-21.2$ and have fluences ranging between $10.8-77.1$\,Jy\,ms, where $7$~bursts have a fluence above the completeness threshold of Westerbork at $24.4$\,Jy\,ms. 

We speculate that the bursts were missed due to a combination of the RFI environment at Westerbork and our large maximum boxcar width of $1024$ which influences the RMS computation. Further tests indicated that a smaller maximum boxcar width ($\sim128$) allowed \texttt{Heimdall} to detect the additional bursts also with the new default S/N-computation method. A detailed investigation of this effect, however, is beyond the scope of this work. 
Motivated by the recovery of these newly identified bursts, while still recovering all previously reported events, we updated our \texttt{Heimdall} configuration to use the original S/N estimation method via the \texttt{boxcar-renorm} option. Our burst searches at Onsala and \torun were not affected, as they use a version of \texttt{Heimdall} built prior to this update. In addition, searches at Stockert and Dwingeloo are performed using \texttt{single\_pulse\_search.py} from the \texttt{PRESTO} software suite. 

\end{document}